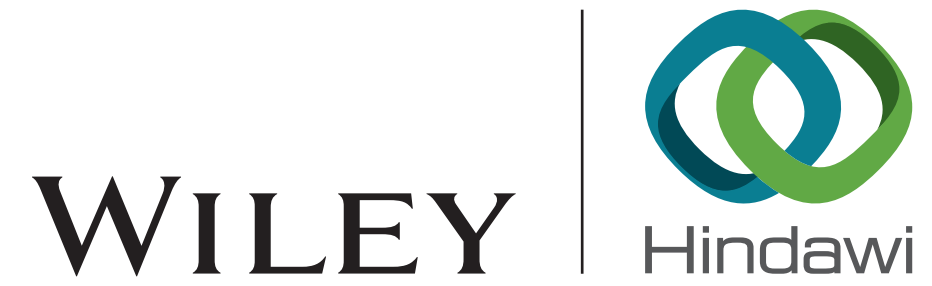

*Research Article*

# Exploration and Coordination of Complementary Multirobot Teams in a Hunter-and-Gatherer Scenario

**Mehdi Dadvar, Saeed Moazami, Harley R. Myler, and Hassan Zargarzadeh**

*Phillip M. Drayer Electrical Engineering Department of Lamar University, Beaumont, TX 77710, USA*

Correspondence should be addressed to Hassan Zargarzadeh; hzargarzadeh@lamar.edu

The hunter-and-gatherer approach copes with the problem of dynamic multirobot task allocation, where tasks are unknowingly distributed over an environment. This approach employs two complementary teams of agents: one agile in exploring (hunters) and another dexterous in completing (gatherers) the tasks. Although this approach has been studied from the task planning point of view in our previous works, the multirobot exploration and coordination aspects of the problem remain uninvestigated. This paper proposes a multirobot exploration algorithm for hunters based on innovative notions of "expected information gain" to minimize the collective cost of task accomplishments in a distributed manner. Besides, we present a coordination solution between hunters and gatherers by integrating the novel notion of profit margins into the concept of expected information gain. Statistical analysis of extensive simulation results confirms the efficacy of the proposed algorithms compared in different environments with varying levels of obstacle complexities. We also demonstrate that the lack of effective coordination between hunters and gatherers significantly distorts the total effectiveness of the planning, especially in environments containing dense obstacles and confined corridors. Finally, it is statistically proven that the overall workload is distributed equally for each type of agent which ensures that the proposed solution is not biased to a particular agent and all agents behave analogously under similar characteristics.

## 1. Introduction

Multirobot systems are expected to complete tasks that are infeasible, laborious, or inefficient for a single agent to accomplish [1]. Employing multirobot systems entails addressing various problems on the subjects of task allocation [2], exploration [3], coordination [4], learning [5], swarm behavior [6, 7], and heterogeneity [8]. Among all of these problems, the problem of multirobot task allocation (MRTA), that is assigning a group of tasks to individual robots, is the most deep-seated problem where its complexity increases considerably in dynamic environments. Since in dynamic problems tasks are unknowingly distributed over an environment, the MRTA problem needs to be addressed from both task planning and multirobot exploration perspectives. The former has been addressed as the hunter-and-gatherer approach in our previous works [9, 10] by dividing each task into two sequential subtasks, where each subtask can only be carried out by a certain type of agent. This novel approach poses an unexplored MRTA problem whose exploration and coordination in complementary teams are the motivation of this work.

According to the taxonomy presented in [11], problems with single-robot (ST) tasks, in which each task requires the effort of a single robot to be completed, are the most primitive cases of MRTA. For instance, the work in [12] addresses MRTA to coordinate a group of autonomous vehicles by proposing two distributed algorithms based on auction and bundle methods. However, in real-world problems, there are cases where each task requires efforts of multiple robots to be completed. This case taxonomically is known as a multirobot (MT) task problem and is investigated in [13, 14]. The former proposes a distributed bees algorithm (DBA) and applies the optimized DBA to distributed target allocation in swarms of robots. The latter presents a novel weighted synergy graph model and then introduces a learning algorithm for the presented model in which the system learns agents' interactions. In both cases, the tasks have been assigned instantaneously, i.e., it is assumed that the tasks are identifiable for robots before the mission. Nonetheless, in a dynamic environment, in which tasks are unknowingly distributed over the environment, instantaneous assignment (IA) is infeasible and instead time-extended assignment (TA) must be dragged in.

In the context of TA, there are mainly two paradigms of works addressing the dynamic problems where tasks are unknowingly distributed over an environment: (1) works that address the problem purely from exploration perspective and (2) works that address the problem from MRTA point of view. Regarding the first paradigm, the authors of [15, 16] present a very fundamental frontier-based algorithm for a single autonomous robot and multi-robot exploration, respectively. To enhance the efficacy of the frontier-based exploration algorithm, Zlot et al. [17] further developed the frontier-based exploration method by introducing a market-based approach to maximize information gain while minimizing incurred costs. Utilizing the theory of information gain in [17] opened the floor to integrate the concept of entropy into the multirobot exploration algorithms. For instance, Bhattacharya et al. [18, 19] are more focused on information theory and cast the exploration problem as minimization of map entropy by taking into account communication among robots. In contrast to [16–19] that consider the whole environment for exploration purposes, Lopez-Perez et al. [20] proposed an algorithm for distributed multirobot system to explore nearby zones to reduce the traversed distance, while agents are efficiently using the resources to communicate with each other. Although [16–20] cope with the unknown nature of the dynamic environments by introducing various multirobot exploration methods, they all neglect integrating the MRTA solution into the proposed exploration algorithms.

Works that fall into the second paradigm undertake environments comprising unknowingly distributed tasks, while addressing the MRTA aspect of the problem. On this subject, Prorok et al. [21] considered a TA problem where a system of heterogeneous robots is modeled as a community of species and developed centralized as well as decentralized methods to efficiently control the heterogeneous swarm of robots. In another effort, in [22], a novel task allocation method is developed based on Gini coefficient which increases the number of accomplished tasks considering limited energy resources. Although Prorok et al. and Wu et al. [21, 22] address a time-extended assignment problem, a solution for exploring the environment to detect unknown tasks has not been provided. Although a few works such as [23] have tried to investigate the performance of task allocation algorithms in a frontier-based multirobot exploration problem, most studies have neglected the integration of multirobot exploration into dynamic MRTA problems such as ST-MR-TA : SP or MT-MR-TA : SP [24]. Consequently, to the best of the authors' knowledge, there is a lack of critical attention paid to addressing multirobot exploration and task allocation simultaneously in TA problems, while this problem is a pervasive problem in a wide variety of fields such as urban search and rescue (USAR) [25], agricultural field operations [26], and security patrols [27].

Aside from the two paradigms reviewed above, foraging is another research trend in addressing multirobot task allocation problem [28]. Although the foraging problem taxonomically falls into the discussed categories in the literature review [11, 24], here we briefly compare the approaches in this paradigm with the hunter-and-gatherer framework. Taken as a whole, foraging is more concerned with collective and swarm behavior of a multiagent system, spanning from ant or bee colony optimization algorithms [29] to multiagent reinforcement approaches [30]. The idea of collective behavior emphasized in foraging research requires identical decision-making mechanism for all agents and results in a decision-making dependency among agents [31]. For instance, it is theoretically challenging to employ an explorer agent with unique search algorithm in the central place foraging algorithm [32] since this exclusive search behavior interferes the swarm behavior of agents and disturbs the system's equilibrium due to decision-making dependency among agents. By way of contrast, solutions to the hunter-and-gatherer framework [9] provide a generic platform for dynamic multiagent task allocation that is more focused on individual autonomy with no decision-making dependency among agents.

Consider the USAR in a disaster site in which a number of victims are stranded in unknown locations and need immediate rescue operations. Each victim is a task that needs to be detected first and then rescued by a rescue operation that typically needs several dexterity actions. This case exemplifies problems where multirobot exploration and task allocation aspects need to be addressed simultaneously. Besides, a rescue robot needs to have a heavy-duty manipulator and dexterous gripper [33, 34], high-power actuators, tracked locomotion mechanism, high-capacity batteries, and many sorts of sensors, cameras, and communication devices to accomplish those tasks which make the robot relatively heavy, ponderous, and incapable of agile search operations. Under this circumstance, the "hunter-and-gatherer approach" is decidedly justifiable, where each task is comprised of two sequential subtasks: detection and completion. Having said that, each subtask can only be carried out by a certain type of agent, where two teams of robots are employed: a team of agile robots that can quickly explore an environment and detect tasks, called "hunters," and a team of dexterous robots who accomplish detected tasks called "gatherers." Practically speaking, hunters can be a group of small UAVs which search the site to locate victims, and gatherers can be a group of maxi-sized [35] heavy-duty UGVs that rescue detected victims relying on their dexterity capabilities.

This paper motivated by the problem explained above, which is taxonomically referred to as ST-MR-TA : SP or MT-MR-TA : SP [24], addresses the dynamic MRTA problem in unknown environments by proposing an integrated multirobot task allocation and exploration solution. According to the hunter-and-gatherer scheme, we first present an innovative decision-making mechanism based on the novel notion of expected gain (EG), which measures density of available information in the surrounding of a potential job (task/frontier). The EG measurement has been integrated into the concept of certainty and uncertainty profit margins by which the levels of agent's confidence and conservativeness are modeled. This innovative decision-making mechanism shapes the background theory of both proposed multi-robot exploration and task allocation algorithms. Besides, this work introduces a coordination factor designated for gatherers through which their behaviors range from completely indifferent to highly coordinated to hunters' locations in the environment. By the way of extensive simulations, we demonstrate that the effectiveness of the proposed algorithms is superior to the performance of the benchmark work [9]. Moreover, statistical analysis of the simulation results shows that the lack of an effective coordination between hunters and gatherers significantly hurts the total effectiveness of the planning. Finally, it is statistically proven that the overall workload is distributed equally for each type of agent which ensures that the proposed solution is not biased to an agent and all agents behave analogously under similar characteristics.

The remainder of this paper is organized as follows: the problem statement is presented in Section 2. In Section 3, the methodology and planning algorithms are discussed. Simulation results are presented in Section 4 followed by conclusion remarks in Section 5.

## 2. Problem Statement

In this section, we present the problem formulation of the hunter-and-gatherer scheme in the context of dynamic MRTA. Assume that there are $m$ tasks distributed randomly over the environment, $E$. We consider a case that the number and the locations of tasks are unknown for agents before the execution of the planning algorithms called hunter-and-gatherer mission planning (HGMP). The set of tasks is denoted as $\mathcal{T} = \{T_1, \ldots, T_m\}$ in which each task is split into hunting and gathering subtasks, i.e., $T_k = \{t_k^h, t_k^g\}$ with $1 \le k \le m$, where $t_k^h$ and $t_k^g$ represent hunting and gathering subtasks, respectively. In this case, the set of agents is defined as $\mathcal{A} = \{A_h, A_g\}$ that comprises of two teams of hunters $A_h = \{a_i^h\}$ and gatherers $A_g = \{a_j^g\}$, where $1 \le i \le n_h$ and $1 \le j \le n_g$. The cost associated with $a_i^h$ for accomplishment of $t_k^h$ is denoted as $c_{k,i}^h$ and $c_{k,j}^g$ is the cost associated with $a_j^g$ for accomplishment of $t_k^g$.

Assumptions: throughout the paper, it is assumed that

(1) Tasks are stationary, i.e., they are fixed to their locations.

(2) The cost of accomplishment of each task is linearly proportional to the distance that an agent moves to do a task. An agent is considered done with a task when it reaches to the task's location.

(3) All agents of a same team are identical.

(4) All agents are rational, i.e., they intend to maximize their expected utility.

(5) All agents are fully autonomous and have their own utility functions, i.e., no global utility function there exists.

(6) Agents from complementary teams can communicate with each other using a stably connected network.

(7) Each location of the grid map is large enough to host multiple agents simultaneously.

(8) Agents autonomously avoid collisions while navigating in a specific location of the grid map simultaneously.

Now, the HGMP problem can be stated as follows. Suppose that there exists a tuple for the mission such that $\text{HGMP} = (E, \alpha, T)$. $\Pi$ denotes the assignment function which assigns $m$ tasks to $n = n_h + n_g$ agents such that $\Pi\colon \mathcal{T} \mapsto \mathcal{A}$. Under the assumptions 1–6, the global objective $\Theta$ is to minimize the collective cost of $\Pi$:

$$\Theta = \min_{x_k^i, y_k^j} \left\{ \rho_h \sum_{i=1}^{n_h} \sum_{k=1}^{m} c_{k,i}^h x_k^i + \rho_g \sum_{j=1}^{n_g} \sum_{k=1}^{m} c_{k,j}^g y_k^j \right\}, \tag{1}$$

where $x_k^i$ and $y_k^j$ are binary decision variables for $t_k^h$ and $t_k^g$:

$$\begin{aligned} x_k^i &\in \{1, 0\}, \quad \forall i, k, \\ y_k^j &\in \{1, 0\}, \quad \forall j, k. \end{aligned} \tag{2}$$

In (1), weighting parameters $\rho_h$ and $\rho_g$ are introduced to sum relative collective costs of complementary teams, because of the physical differences of each type.

This problem has a global objective $\Theta$ which can be achieved by determining the binary decision variables optimally. However, finding the optimal solution in multirobot path planning and multirobot task planning problems is NP-hard, as proven in [36–38], respectively. That being said, addressing such problems from agents' point of view is an admissible approach to find relatively better solutions, i.e., local optimal solutions. Since agents are rational, each agent's objective is to maximize its own expected utility in distributed approaches. Therefore, the aim of this paper is to design a distributed decision-making mechanism which allows agents to maximize their own expected utility while the individual efforts converge to a local optimal solution from the society standpoint. In other words, the binary decision variables in $\Theta$ need to be determined by the agents throughout explorations and coordination in a distributed manner. This approach also necessitates a study on the hyperparameters of the proposed algorithm to demonstrate how local optimal solutions are achieved by adjusting those parameters with respect to the practical admissible ranges of parameters in each scenario.

Besides, in the proposed problem statement, the allocation problem is considered dynamic for multiple reasons. First, tasks are unknowingly distributed over the environment. Thus, agents do not have any prior information regarding the tasks' location and need to explore the environment to identify them. Secondly, based on the problem statement, there are always $m$ tasks in the environment, i.e., when a task is accomplished by the agents, another task will be distributed randomly over the environment. Altogether, it is not feasible theoretically and practically to accomplish the planning right after the start of a mission. Instead, only dynamic planning algorithms can cope with the unknown and dynamic nature of the environment.

## 3. Methodology

*3.1. Conceptual Frameworks.* Hunters are assigned to explore the unknown environment for detecting new tasks. According to the hunter-and-gatherer scheme, the detected task can only be completed by a gatherer's effort. Since we aim to develop the planning algorithms in a distributed manner, there should be stably connected communication between agents from complementary teams. Considering that fact, hunters announce the location of any newly detected tasks so that gatherers can decide about accomplishing them. Since there is no peer-to-peer communication and all communications are supposed to be broadcasted, we name the communication platform an "online board" through which gatherers notice the location of new detections.

In this section, we develop reasoning mechanisms for both types of agents to properly achieve the global objective of this work mentioned in Section 2. We first illustrate the concept of certainty and uncertainty profit margins, which are the building blocks of the reasoning mechanism of both types of agents. Secondly, we propose a multirobot exploration algorithm for hunters in a distributed manner by introducing the notion of EG incorporated into the concept of profit margins. Subsequently, the way that gatherers accomplish detected tasks is delineated based on the same theoretical frameworks. In fact, we elucidate how the same theories of profit margins and EG can be generalized to develop the multirobot task planning and coordination algorithm of gatherers.

*3.2. Notion of Profit Margins.* The rationale behind the idea of profit margins is to classify potential jobs (tasks/frontiers) in an environment into profitable, weakly-profitable, and nonprofitable types. When a job is profitable, the agent is confident about taking actions to accomplish it. On the other hand, the agent is conservative about potential jobs that are weakly-profitable and ignores nonprofitable jobs. The effort needed to accomplish a job is the factor that determines whether a job is profitable, weakly-profitable, or nonprofitable. According to the second assumption, the effort made by an agent to accomplish a job corresponds to the distance that it travels to reach the job. For example, the effort that a gatherer makes to accomplish a job is the distance that it travels to reach and accomplish a task. Similarly, the effort that a hunter makes to accomplish its job is the distance that it travels to explore the environment by reaching the frontiers.

Now, we define the certainty and uncertainty profit margins (CPM and UPM) more specifically for both types of agents with respect to the accomplishment cost of a job. CPM is a margin to which the travel distance is less than $R_c$ from agent's perspective. UPM is a margin to which the travel distance is less that $R_u$ and greater than $R_c$ from agent's point of view. Figure 1 shows the CPM and UPM conceptually as two concentric circles with the agent at the center. In the case of this figure, Job 1 is included in agent's CPM, so it is considered as a profitable job and the agent is confident to accomplish it. Further, the agent is conservative about completing Job 2 since it falls in its UPM and is a weakly-profitable job. Finally, Job 3 is located beyond the agent's UPM, so it is not profitable, and the agent ignores it.

Since agents of both type function in an environment in the presence of obstacles, we explain the concept of CMP and UMP for an agent functioning in an occupancy grid map [39]. Figure 2 illustrates an occupancy grid map with an agent located at the center. In this figure, the concept of profit margins has been applied to the probabilistic road maps (PRM) generated for agent's path planning. In other words, Figure 2 explains how an agent practically classifies jobs as profitable, weakly profitable, and nonportable in a map relying on the PRMs.

$$\varepsilon_{c,z}^{f} = \frac{\lambda_z^c}{d_{c,z}^{f} \sum_{p=1}^{\lambda_z^c} \alpha_p^c}. \tag{3}$$

According to the hunter-and gatherer scheme, a hunter agent relies to its profit margins to explore the environment and a gatherer agent considers its profit margins to accomplish detected tasks. Regarding the definition of UPM and CPM and the way that it can be applied to PRMs, we focus on developing the reasoning mechanisms for both types of hunter-and-gatherer agents in the subsequent sections.

*3.3. Reasoning Mechanism: Hunters.* In this section, we aim to develop a reasoning mechanism based on the definition of profit margins so that hunters explore the environment. In this regard, we utilize the frontier-based exploration concept to develop a CPM and UPM-based multirobot exploration algorithm. The basic idea in a frontier-based exploration algorithm is that the explorer agent selects a frontier point first and then moves towards the selected frontier to explore unknown areas iteratively. Although we develop the reasoning mechanism for hunters in a distributed manner, we need to utilize an online shared map in which certain information of the map and frontiers are accessible for all agents. Hence, before developing the reasoning mechanism, we define a platform in which agents share their gained information.

We define an online board which contains the collective gained information about the environment's map. At the beginning of each mission, all cells of the occupancy grid map are marked as unknown. While hunters explore the map, each explored cell can be marked as obstacle (cells with probability greater than 0.5 in the occupancy grid map), free, or task cell. Moreover, the unknown cells neighboring a known cell will be marked as a frontier cell. By analyzing the data embedded in the online board, each hunter decides which frontier to select and explore in a distributed manner relying its reasoning mechanism. The reasoning mechanism splits into two steps: (1) the map updating process, i.e., the hunter updates some additional information on each frontier cell collectively, and (2) the decision process, i.e., the process by which the hunter chooses a frontier cell to explore.

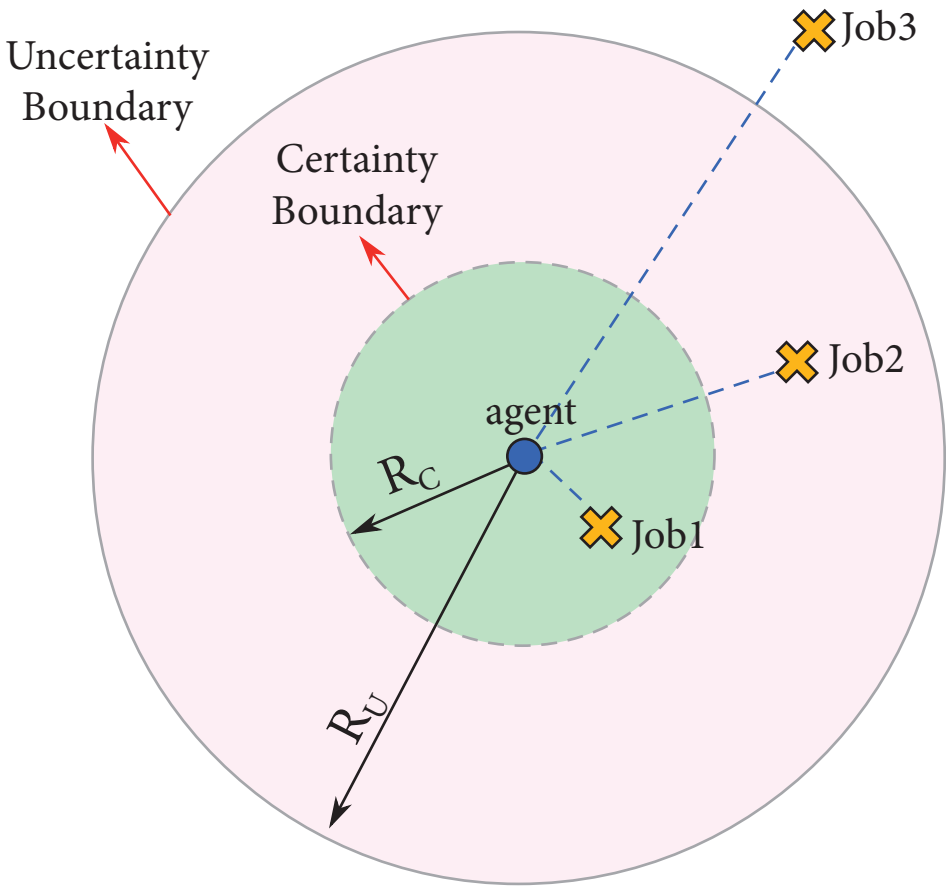

Figure 1: CPM and UPM of an agent: Job 1 and Job 2 are in the agent's CPM and UPM, respectively. Job 3 is beyond the agent's uncertainty boundary.

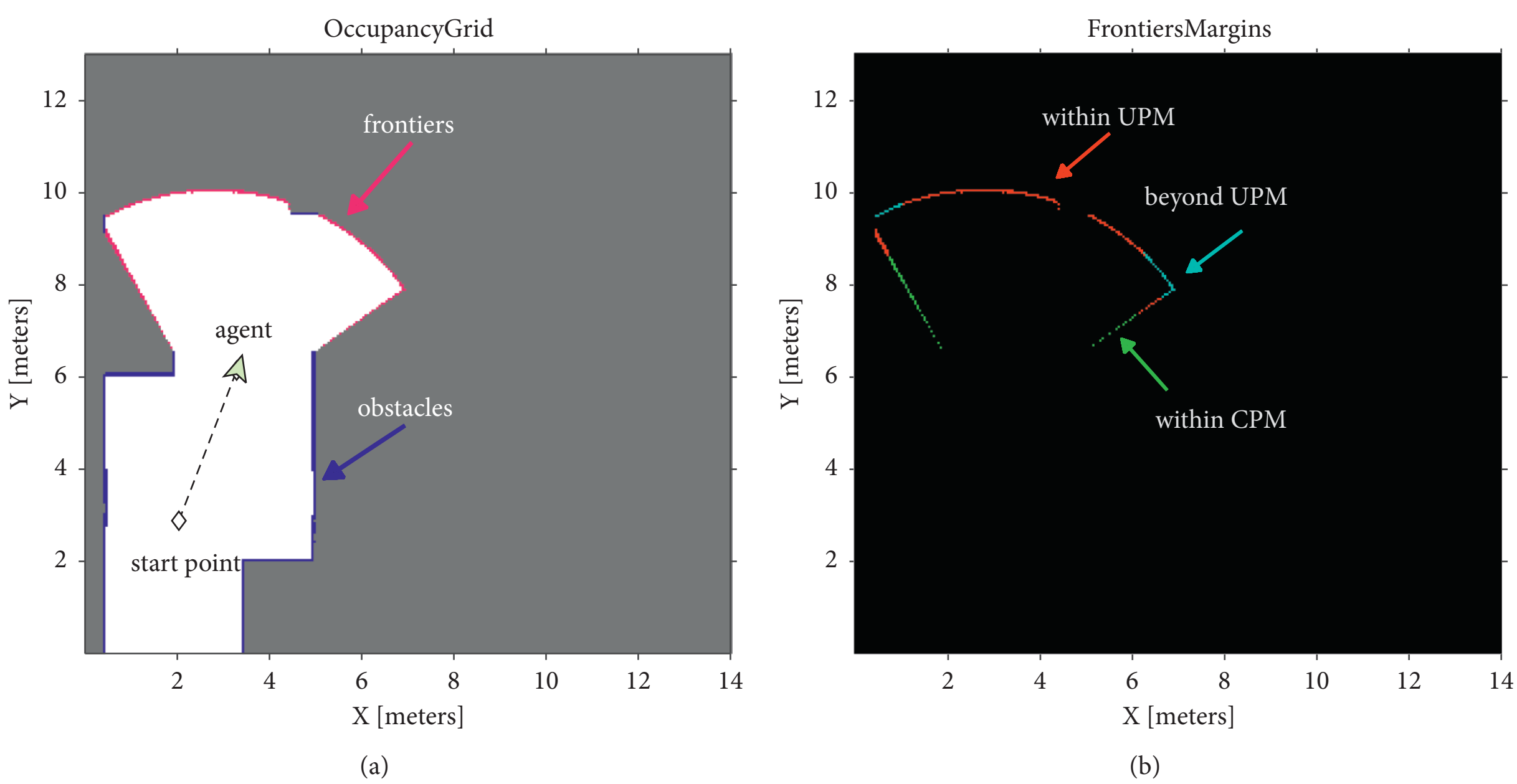

Figure 2: Frontiers in the agent's occupancy grid map: (a) the updated map from an agent's point of view with presence of obstacles and detected frontiers, and (b) categorization of frontiers according to the definition of CPM and UPM.

Regarding the first step, the hunter agent classifies all frontiers into three categories according to the definition of CPM and UPM, as illustrated in Figure 3. Then, the hunter updates the location of detected frontiers on the online board and the hunter agent indicates that if the new frontiers fall within its CPM or UPM. To elaborate, each frontier cell keeps two factors called certainty and uncertainty factors (CF and UF). CF of a frontier indicates the number of hunters that the frontier cell is included in their CPM. Similarly, UF of a frontier cell indicates the number of hunters that the frontier cell is within their UPM. Accordingly, the hunter updates CF and UF of all frontiers within its CPM and UPM. In each iteration, the hunter does the map updating process first and then relies on the CF and UF information of frontiers to proceed the decision process.

As elaborated above, we need to develop a decision process by which a hunter decides which frontier to choose for exploring relying on the information updated on the online board. Here, we propose a method which considers the EG available by exploring a certain frontier and chooses a frontier with maximum value of EG. This method has three main features: (1) the algorithmic method is developed in a distributed manner, so we propose the decision process for an instance hunter agent, (2) the relative position of other hunters is being considered in the decision process (using CFs and UFs) which predictably prevents hunters from rushing towards closely similar regions, and (3) the CFs and UFs of all neighbor frontiers are reflected in defining EG for a candidate frontier to guarantee the previous property. As a matter of fact, the neighborhood of a frontier corresponds to the CPM of that frontier.

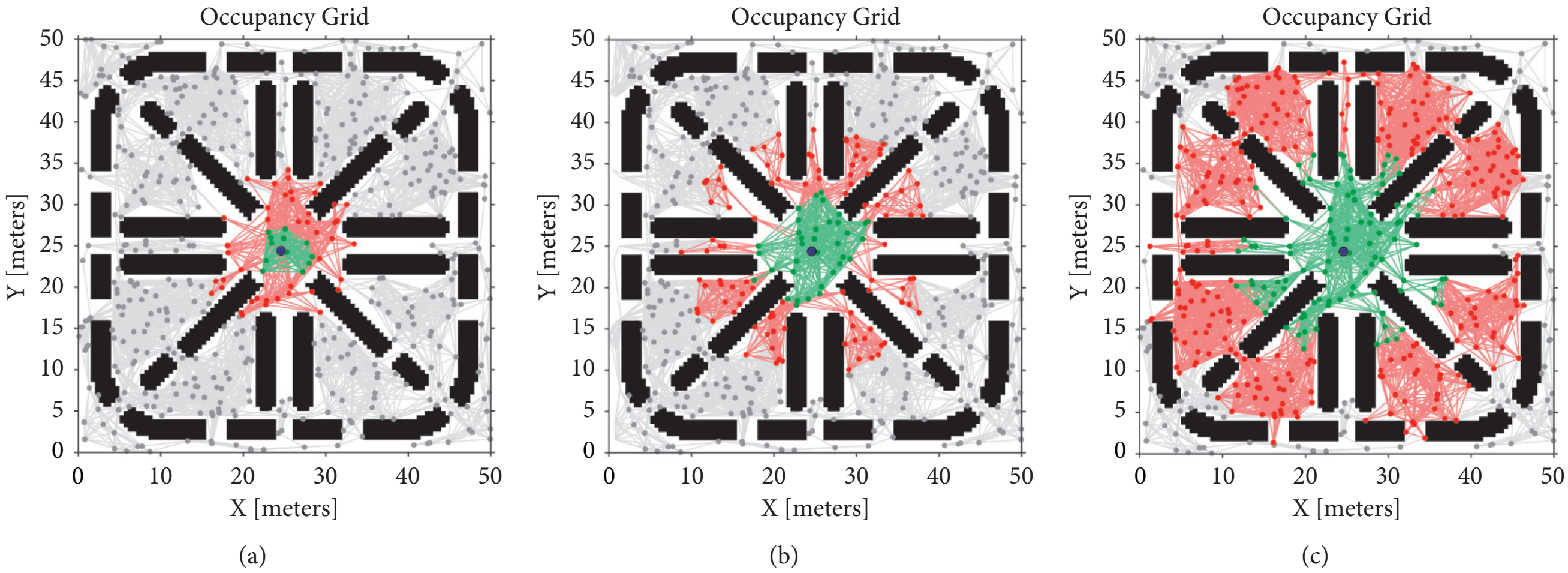

FIGURE 3: This figure explains CPM and UPM in an occupancy grid map with presence of obstacles, where the agent is located at the center. In the generated PRM, gray nodes represent nodes that are beyond agent's profit margins, while green and red nodes represent nodes which are within the agent's CPM and UPM, respectively. CPM and UPM have been calculated for different values of Rcand Ru: (a) Rc Ł 4 mand Ru Ł 10 m, (b) Rc Ł 8 m and Ru Ł 16 m, and (c) Rc Ł 14 mand Ru Ł 25 m

As explained for the map updating process, the hunter classifies all frontiers into 3 classes regarding its CPM and UPM. In this step, we clarify how EG is defined for frontiers within agent's CPM. Afterwards, we will develop the EG definition for frontiers within hunter's UPM. Needless to mention, frontiers beyond hunter's UPM are ignored by the agent due to the definition of profit margins.

Suppose that there are $\lambda^c$ frontiers within the hunter's CPM where $\lambda^c \geq 1$. Then, the set of frontiers within its CPM is defined as $F^c = \{f_z^c\}$, where $1 \leq z \leq \lambda^c$. The hunter, denoted as $a_i^h$, needs to calculate EG for all members of $F^c$ and then choose a frontier with the highest value of EG. A primary factor which effects EG of a frontier is the distance between the hunter and the frontier such that EG $\propto$ distance$^{-1}$. This proportionality needs to be completed by considering other conditions of the frontier to have a more accurate definition of EG. Let $f_z^c$ denotes the candidate frontier which $a_i^h$ aims to analyze and calculate its EG $a_i^h$ needs to know if it visits $f_z^c$, then the agent determines how many other frontiers are available within the CPM (neighborhood) and what is the collective CF of those frontiers. Figure 4(a) illustrates an example in which a hunter agent has already classified all frontiers available on the online board. Initially, the hunter chooses $f_z^c$ as a candidate frontier among all frontiers which are within the hunter's CPM. In addition, Figure 4(b) explains how the hunter determines the frontiers within the CPM of the candidate frontier $f_z^c$. Let $\lambda_z^c$ denote the number of frontiers within CPM of $f_z^c$. Accordingly, the set of expected frontiers with respect to $f_z^c$ is defined as $f_z^{*c} = \{f_{z,p}^c\}$, where $1 \leq p \leq \lambda_z^c$. Next, $a_i^h$ calculates the collective CF of all members in $f_z^{*c}$. Now, EG $\propto$ distance$^{-1}$ gets completed by adding a coefficient which is the ratio of $\lambda_z^c$ and the collective CFs of $f_z^{*c}$. The reason that we consider only CF is that the candidate frontier itself is within agent's CPM. The set of EG for all frontiers within agent's CPM is denoted as $\varepsilon_c^f = \{\varepsilon_{c,z}^f\}$, where $1 \leq z \leq \lambda^c$, and $\varepsilon_{c,z}^f$ denotes the EG of a candidate frontier, i.e., $f_z^c$. Further, CF of a member of $f_z^{*c}$ is denoted as $\alpha_p^c$. Finally, $\varepsilon_{c,z}^f$ for a candidate frontier within agent's CPM is defined as follows:

To put it simply, we have EG $\propto$ distance$^{-1}$ for each frontier. Then, distance$^{-1}$ is multiplied by a coefficient $(\lambda_z^c / \sum_{p=1}^{\lambda_z^c} \alpha_p^c)$ in which its numerator is the total number of frontiers available in the candidate frontier's neighborhood and its denominator is the collective CFs of those frontiers. In other words, higher values of the numerator indicate that there are other frontiers in the candidate frontier's neighborhood which can be accessible for the agent to explore easily when it visits it. However, the denominator reflects the presence of other hunters within the candidate frontier's neighborhood.

Altogether, $a_i^h$ calculates $\varepsilon_{c,z}^f$ for all frontiers within its CPM and then chooses a frontier with the maximum value of expected information gain, denoted as $f_\zeta^c$, such that

$$\zeta^f = \text{argsmax}\left(\varepsilon_c^f\right). \tag{4}$$

Similarly, EG can be defined for a candidate frontier within hunter's UPM with a slight difference. In this case, both collective CFs and UFs will be considered to define EG for a candidate frontier. To explain, suppose there are $\lambda^u$ frontiers within the hunter's UPM where $\lambda^u \geq 1$ and $\lambda^c = 0$. Then, the set of frontiers within a candidate frontier's UPM is defined as $F^u = \{f_z^u\}$ where $1 \leq z \leq \lambda^u$. $a_i^h$ needs to calculate EG for all members of $F^u$ and chooses a frontier with the highest value of EG. Let $f_z^u$ denote the candidate frontier in which $a_i^h$ aims to analyze and calculate its EG. The hunter $a_i^h$ needs to know that when it visits $f_z^u$, then how many frontiers are available within its CPM, and what are the collective CF and UF of those frontiers. Let $\lambda_z^u$ denote the number of frontiers within CPM of $f_z^u$. Accordingly, the set of expected frontiers with respect to $f_z^u$ is defined as $f_z^{*u} = \left\{f_{z,p}^u\right\}$, where $1 \le p \le \lambda_z^u$. Next, $a_i^h$ calculates the collective CF and UF for all members of $f_z^{*u}$ by considering the online board. Since $f_z^u$ is located within UPM of $a_i^h$, then it considers both CF and UF to calculate EIG. The set of EG for all frontiers within agent's UPM is denoted as $\varepsilon_u^f = \left\{\varepsilon_{u,z}^f\right\}$, where $1 \le z \le \lambda^u$, and $\varepsilon_{u,z}^f$ denotes the EG of a candidate frontier, i.e, $f_z^u$. Further, CF and UF of a member of $f_z^{*u}$ are denoted as $\alpha_p^u$ and $\beta_p^u$, respectively. Finally, $\varepsilon_{u,z}^f$ for a candidate frontier within agent's UPM is defined as

$$\varepsilon_{u,z}^f = \frac{\lambda_z^u}{d_{u,z}^f\left(\sum_{p=1}^{\lambda_z^u} \alpha_p^u + \sum_{p=1}^{\lambda_z^u} \beta_p^u\right)}, \quad (5)$$

where $a_i^h$ calculates $\varepsilon_{u,z}^f$ for all frontiers within its UPM and then chooses a frontier with the maximum value of EG, denoted as $f_\zeta^u$, such that

$$\zeta^f = \operatorname{argsmax}\left(\varepsilon_u^f\right). \quad (6)$$

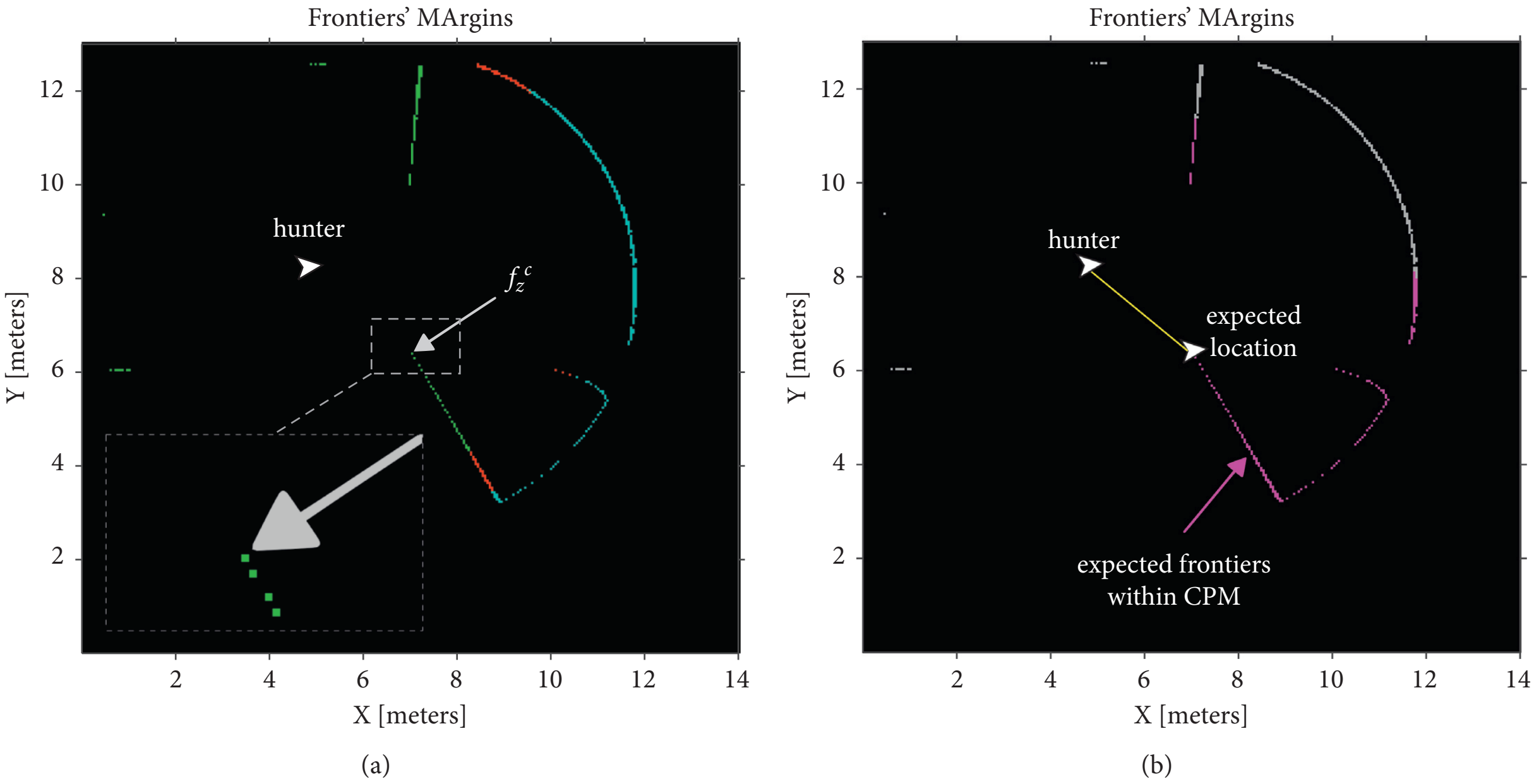

Figure 4: An example illustrating the way a hunter calculates the number and collective CF of frontiers within a candidate frontier's CPM: (a) categorization of frontiers and selecting a candidate frontier, and (b) the frontiers available within the candidate frontier's CMP are in purple. In other words, if the hunter visits the candidate frontier, then all purple frontiers will be accessible within its CPM (neighborhood).

The above procedures for selecting a frontier have been considered to develop a frontier selection function. Algorithm 1 illustrates the procedure in which a hunter selects a frontier within its CPM or UPM. In line 3, the hunter uses the definition of profit margins, i.e., $R_c$ and $R_u$, to categorize all frontiers and updates the CF and UF of frontiers on the online board. In lines 5 and 9, the agent utilizes (3) and (5) respectively to calculate EGs. Further, the hunter uses (4) and (6) to choose a frontier with highest value of EG in lines 7 and 9, respectively.

The frontier selection function explained is Algorithm 1 needs to be invoked in the hunter's main algorithm. To that end, Algorithm 2 illustrates the main decision procedure for a hunter agent. In line 1, $\tau_{\max}$ denotes the maximum number of iterations at which the main procedure is executed. In line 2, the hunter checks to know whether its frontier buffer is empty to invoke the frontier selection function. In line 4, the agent updates the status of the selected frontier on the online board. In fact, since the online board stores the location of all agents and also the frontier map of the environment, the CF and UF of each frontier are calculated in a centralized manner and are available on it. When the selected frontier is located within agent's CPM, then after updating, the frontier is not selectable for other agents. Otherwise, the agent only updates the status of the selected frontier to pending which still allows other agents, i.e., agents that the selected frontier is within their CPM, to select the frontier. In other words, when the selected frontier is within agent's UPM, then there are still chances for other closer agents to select the frontier. This is a reassignment process which results in improving the assignment iteratively regarding the dynamics of the environment. However, when the agent gets close enough to the selected frontier so that the frontier becomes included in its CPM, the agent can update the status of the selected frontier such that no reassignment be allowed anymore. In line 7, the agent checks the condition to make sure whether the selected frontier is still available. Obviously, when the agent selects a frontier within its CPM, then this condition is always true. When a frontier is selected and is still available, then the hunter iteratively moves towards the selected frontier. In line 9, relying on the sensor data, the hunter

1: **function** ChooseFrontier ($R_c$, $R_u$, online board)

2: **for** all detected frontiers in online board **do**

3: $F = \{F^c, F^u\} \longleftarrow$ categorize frontiers

4: **end for**

5: **if** $F^c \neq \varnothing$, **then**

6: $\varepsilon_c^f \longleftarrow$ Calculate EG set

7: $f_\zeta^c \longleftarrow$ choose the frontier with the highest EG

8: **return** $f_\zeta^c$

9: **else if** $F^c = \varnothing$ and $F^u \neq \varnothing$, **then**

10: $\varepsilon_u^f \longleftarrow$ Calculate EG set

11: $f_\zeta^u \longleftarrow$ choose the frontier with the highest EG

12: **return** $f_\zeta^u$

13: **else**

14: **return** $\varnothing$

15: **end if**

16: **end function**

Algorithm 1: Frontier selection function.

1: **for** $\tau = 1{:}\ \tau_{\max}$, **do**

2: **if** frontierBuffer = $\varnothing$, **then**

3: frontierBuffer $\longleftarrow$ ChooseFrontier ($R_c$, $R_u$, OB)

4: **end if**

5: Update the status of the selected frontier on OB

6: **if** frontierBuffer $\neq \varnothing$, **then**

7: **if** the selected frontier is still available, **then**

8: move towards the selected frontier

9: **if** a new task is detected, **then**

10: update OB

11: **end if**

12: **else**

13: frontierBuffer $\longleftarrow \varnothing$

14: **end if**

15: **end if**

16: Update newly detected frontiers on OB

17: **end for**

Algorithm 2: A hunter agent's iterative main loop.

checks whether a new task is detected while moving towards the selected frontier. In line 16, the hunter updates the new detected frontiers on the online board according to the updated captured data, while moving towards the selected frontier. To clarify, when a selected frontier is within the agent's CPM, then the agent is responsible for exploring the corresponding area of the selected frontier, which makes it impossible for other agents to select that frontier.

*3.4. Reasoning Mechanism: Gatherers.* In this section, we aim to develop a reasoning mechanism based on profit margins so that gatherers accomplish the detected tasks efficiently. On this subject, we develop a task-selection algorithm like the frontier selection algorithm in the previous section, but we also consider the coordination between gatherers and hunters to develop the algorithm. To this end, the EG for a task is a function of locations of tasks and frontiers. In fact, the locations of tasks play the main role to calculate the EG, but the locations of frontiers are also considered to involve the coordination factor between a gatherer and the other hunters. This effectively enables a gatherer agent to prioritize tasks surrounded by more frontiers because any region with higher density of frontiers is more susceptible for the presence of hunters. This reasoning rationally performs a coordination between gatherers, which are accomplishing detected tasks, and hunters, which are exploring the environment by visiting the frontiers.

The reasoning mechanism for a gatherer agent splits into two steps: (1) the map updating process, i.e., the gatherer updates some additional information on each task-marked cell, and (2) the decision process, i.e., the process by which the gatherer chooses a task to accomplish. To do the map updating process, the gatherer agent classifies all detected tasks into three categories according to the definition of CPM and UPM. Then, the gatherer updates CF and UF factors of each task-marked cell. The CF of a task indicates the number of gatherers that the task is included within their CPM. The UF of a task-marked cell indicates the number of gatherers that the task is within their UPM. Accordingly, the gatherer updates CF and UF of all detected tasks within its CPM and UPM. In each iteration, the gatherer does the map updating process first and then relies on the CF and UF information of detected tasks to proceed the decision process. To develop the decision process for a gatherer agent, we first clarify how EG is defined for tasks within agent's CPM and then we develop the EG definition for tasks within gatherer's UPM.

Suppose that there are $\kappa^c$ tasks within the gatherer's CPM where $\kappa^c \geq 1$. Then, the set of tasks within its CPM is defined as $T^c = \{t_z^c\}$ where $1 \leq z \leq \kappa^c$. The gatherer, denoted as $a_j^g$, needs to calculate EG for all members of $T^c$ and then choose a task with the highest value of EG. A primary factor which effects EG of a task is the distance between the gatherer and the task such that EG $\propto$ distance$^{-1}$. This proportionality needs to be completed by considering other conditions of the task to have a more accurate definition of EG. Let $t_z^c$ denote the candidate task that $a_j^g$ aims to analyze and calculate its EG. $a_j^g$ needs to know if it visits $t_z^c$, then how many other tasks will be available within its CPM, and what is the collective CF of those tasks. Let $\kappa_z^c$ denote the number of tasks within CPM of $t_z^c$. Accordingly, the set of expected tasks with respect to $t_z^c$ is defined as $t_z^{*c} = \{t_{z,p}^c\}$ where $1 \leq p \leq \kappa_z^c$. Next, $a_j^g$ calculates the collective CF of all members in$t_z^{*c}$. Now, EG $\propto$ distance$^{-1}$ gets completed by adding a coefficient which is the ratio of $\kappa_z^c$ and the collective CFs of $t_z^{*c}$. The set of EG for all tasks within agent's CPM is denoted as $\varepsilon_c^t = \{\varepsilon_{c,z}^t\}$ where $1 \leq z \leq \kappa^c$ and $\varepsilon_{c,z}^t$ denotes the EG of a candidate task, i.e., $t_z^c$. Further, CF of a member of $t_z^{*c}$ is denoted as $\sigma_p^c$. Therefore, for $\varepsilon_{c,z}^t$ of a candidate task within agent's CPM, we have

$$\varepsilon_{c,z}^t \propto \frac{\kappa_z^c}{d_{c,z}^t \sum_{p=1}^{\kappa_z^c} \sigma_p^c}. \tag{7}$$

To complete (7), we also need to consider the coordination between the gatherer and other hunters by caring about the availability of frontiers within CPM of the candidate task. To that end, we will multiply the right side of (7) by a coordination term which is $(1 + \mu\lambda_z^c)$, where $\lambda_z^c$ and $\mu$ denote the number of frontiers within CPM of $t_z^c$ and the coordination coefficient, respectively. Thus, $\varepsilon_{c,z}^t$ of a candidate task within agent's CPM is defined as follows:

$$\varepsilon_{c,z}^t = \frac{\kappa_z^c}{d_{c,z}^t \sum_{p=1}^{\kappa_z^c} \sigma_p^c} \left(1 + \mu\lambda_z^c\right). \tag{8}$$

Similarly, $a_j^g$ calculates $\varepsilon_{c,z}^t$ for all tasks within its CPM and then chooses a task with the maximum value of EG such that $\zeta = \text{argsmax}(\varepsilon_c^f)$ where $t_\zeta^c$ denote the chosen task. The admissible values for the coordination coefficient are defined as $\mu \geq 0$. However, the optimal value of it depends on the dimension and floor map of the environment and also depends on the number of hunter agents.

By the same token, EG can be defined for a candidate task within a gatherer's UPM with a slight difference. In this case, both collective CFs and UFs will be considered to define EG for a candidate task. To illustrate, suppose that there are $\kappa^u$ tasks within the gatherer's UPM where $\kappa^u \geq 1$ and$\kappa^c = 0$. Then, the set of tasks within its UPM is defined as $T^u = \{t_z^u\}$ where $1 \leq z \leq \kappa^u$. The gatherer, denoted as $a_j^g$, needs to calculate EG for all members of $T^u$ and then choose a task with the highest value of EG. Let $t_z^u$ denote the candidate task in which $a_j^g$ aims to analyze and calculate its EG. $a_j^g$needs to know if it visits $t_z^u$, then how many tasks will be available within its CPM, and what are the collective CF and UF of those tasks. Let $\kappa_z^u$ denote the number of tasks within CPM of $t_z^u$. Accordingly, the set of expected tasks with respect to $t_z^u$ is defined as $t_z^{*u} = \{t_{z,p}^u\}$ where $1 \leq p \leq \kappa_z^u$. Next, $a_j^g$ calculates the collective CF and UF of all members in$t_z^{*u}$. The set of EG for all tasks within agent's UPM is denoted as $\varepsilon_u^t = \{\varepsilon_{u,z}^t\}$ where $1 \leq z \leq \kappa^u$ and $\varepsilon_{u,z}^t$ denotes the EG of a candidate task, i.e., $t_z^u$. Further, CF and UF of a member of $t_z^{*u}$ are denoted as $\sigma_p^u$ and $\omega_p^u$, respectively. Additionally, $\kappa_z^c$ denotes the number of frontiers within CPM of $t_z^u$. Therefore, for $\varepsilon_{u,z}^t$ of a candidate task within agent's UPM, we have

$$\varepsilon_{u,z}^t = \frac{\kappa_z^u}{d_{u,z}^t \left(\sum_{p=1}^{\kappa_z^u} \sigma_p^u + \sum_{p=1}^{\kappa_z^u} \omega_p^u\right)} \left(1 + \mu\lambda_z^u\right). \tag{9}$$

Similarly, $a_j^g$ calculates $\varepsilon_{u,z}^t$ for all tasks within its UPM and then chooses a task with the maximum value of EG such that $\zeta = \text{argsmax}(\varepsilon_u^f)$ where $t_\zeta^u$ denotes the chosen task.

The task-selection procedure is almost like the procedure illustrated in Algorithms 1 and 2. The main difference is the way that a gatherer calculates EG for all detected tasks which has been illustrated by (8) and (9).

## 4. Simulation Results

In this section, we put the exploration and coordination algorithms in the hunter-and-gatherer scenario into test by running extensive simulations and investigating the performance of the proposed method statistically from multiple aspects. First, we aim to validate the fairness of the proposed task allocation algorithm. This validation, which is carried out by comparing agents' effectiveness in a set of experiments and analyzing the results by paired $T$-test and ANOVA [40] methods, ensures that the overall workload of a mission is distributed equally among agents of both type. Thereafter, we study the effect of profit margins on the total effectiveness of the proposed methods by accomplishing parameter studies on $R_C$ and $R_U$. After that, we need to study the efficacy of the introduced coordination factor for gatherers by investigating its effect on the planning's total effectiveness. As the final steps, the functionality of the proposed method is tried out by drawing comparisons. To that end, we first compare the performance of the proposed method with the benchmark hunter-and-gatherer approach introduced by the authors' previous work in multiple environments, and then the functionality of the exploration and coordination algorithms in the context of the hunter-and-gatherer scheme is verified by a comparison of its performance and a basic alternative method in which each agent does both hunting and gathering tasks itself. These two comparisons ensure that the newly proposed method outperforms the benchmark method while it is superior to the nonhunter-and-gatherer approaches.

To simulate the proposed approaches, we developed a multirobot simulation platform in MATLAB from scratch. In this platform, we can implement the simulations on any custom map, while the number of each type of agent is adjustable. We provide some basic functions for each type of agent to enable them maneuver over the determined environment. For gatherers, we utilized $A^*$-based motion planning algorithm which enables them to move along two points in a grid environment. Besides, the number of tasks is also adjustable while they get located randomly over the environment. As a matter of fact, we also provided the perpetual mode for implantation of the simulations where, for each gathered task, another task will be distributed randomly in the environment. Accordingly, at each iteration, there are certain number of tasks available in the environment which is adjustable for each mission. Further, in the perpetual mode, each explored and known grid of the environment turns into an unknown grid after certain iterations. The perpetual mode helps the analysis be done in a much more accurate and evidence-based way.

All simulations have been executed under the following conditions: (1) the environment is sectioned as a $e_L \times e_W$ grid of tiles where $e_L = e_W = 100$, (2) the quantities of each type of agents are adjusted as $n_h = 4$ and $n_g = 2$, (3) there are always $m_p = 25$ tasks in the environment, (4) the maximum number of iterations is determined as $\tau_{\max} = 1000$, and (5) we considered the weighting parameters as $\rho_h/\rho_g = 0.2$. Moreover, the simulations have been conducted with abstract agents in order to experiment and evaluate the proposed algorithms more generically and without being biased to any specific types of agents.

*4.1. Task Allocation Fairness.* To demonstrate that the accomplishment's workload is distributed equally for each type of agent, the concept of fairness is introduced. We need to investigate the task allocation algorithms from the fairness perspective for two main reasons: (1) to prove that the allocation is not biased to a particular agent by ensuring that agents behave analogously under similar characteristics, and (2) to confirm that there is no imbalance in agent's involvement in a mission which, practically speaking, results in an equal wear and tear of individual robots while operating in real-world situations.

We define an effectiveness factor for each agent of both types based on their costs and accomplishment. Then, using the statistical analysis, the fairness of the HGM by comparing effectiveness of different agents of each type can be proven. Let $\eta_i^h$ and $\gamma_i^h$ denote the effectiveness of $a_i^h$ and the number of tasks hunted by the agent, respectively, as the following:

$$\eta_i^h = \frac{\gamma_i^h}{\sum_{k=1}^{m} c_{k,i}^h x_k^i}. \tag{10}$$

Similarly, $\eta_j^g$ and $\gamma_j^g$ denote the effectiveness of $a_j^g$ and the total number of tasks gathered by the agent, respectively, such that

$$\eta_j^g = \frac{\gamma_j^g}{\sum_{k=1}^{m} c_{k,j}^g y_k^j}. \tag{11}$$

To investigate the fairness of the proposed algorithms, we ran 100 missions and recorded agents' effectiveness according to (10) and (11). Then, utilizing statistical hypothesis testing, we prove the fairness for each type of agents by showing that the averages of agents' effectiveness in 100 missions are statistically identical. Having said that, the hypothesis testing will be applied on the mean of 100, recorded effectiveness for each hunter and gatherer. Figure 5(a) shows the statistical results of $\eta_i^h$ for all hunters. As $n_h = 4$, an ANOVA test has been applied to the collected data to statistically prove the fairness of the proposed algorithms for hunters. The ANOVA test has been applied as follows: $H_0$: $\mu_1^h = \mu_2^h = \mu_3^h = \mu_4^h$, $H_1$: $\mu_1^h \neq \mu_i^h$, and $\alpha = 0.05$, where $\mu_i^h$ denotes the average of $\eta_i^h$ for $a_i^h$ in 100 tests and $\alpha$ denotes the significance level. According to the results of the ANOVA test, we have $F = 0.39$, $F - \text{crit} = 2.62$, and $P - \text{value} = 0.75$. Since $F < F - \text{crit}$ and $P - \text{value} > \alpha$, we must retain the null hypothesis. Thus, it has been proved that $\mu_1^h = \mu_2^h = \mu_3^h = \mu_4^h$. In addition, as $n_g = 2$, a paired $T$-test has been applied to the data to investigate the fairness of the proposed algorithms for gatherers. The hypothesis testing has been done in such a manner $H_0$: $\mu_1^g - \mu_2^g = \mathrm{D}_0$, $H_1$: $\mu_1^g - \mu_2^g \neq \mathrm{D}_0$, $D_0 = 0$, $n_s = 100$, dof $= 99$, and $\alpha = 0.05$. According to the test $p - \text{value} = 0.1$. Since, $p - \text{value} > \alpha$, we must retain the null hypothesis. Therefore, it has been proven that $\mu_1^g - \mu_2^g = D_0 = 0$, as it is illustrated in Figure 5(b).

*4.2. Effect of Agent's Profit Margins on the Total Effectiveness.* The proposed algorithms rely strongly on introduced definitions of profit margins, as discussed in the methodology section. Accordingly, we need to study the effect of profit margins' parameters on a mission's effectiveness to demonstrate their functionality for both types of agents. For this reason, we define the effectiveness for a mission, denoted as $\eta_t$, which is the ratio of the total number of completed tasks, $\gamma_t$, and the collective cost of the whole mission, $C_t$, as follows:

$$C_t = \rho_h \sum_{i=1}^{n_h} \sum_{k=1}^{m} c_{k,i}^h x_k^i + \rho_g \sum_{j=1}^{n_g} \sum_{k=1}^{m} c_{k,j}^g y_k^j, \tag{12}$$

$$\gamma_t = \sum_{j=1}^{n_g} \gamma_j^g, \tag{13}$$

$$\eta_t = \frac{\gamma_T}{C_T}. \tag{14}$$

Regarding (14), extensive simulations have been ran for all values of $R_c^h$ and $R_u^h$, i.e., the profit margins of hunters, such that $1 \leq R_c^h \leq 136$ and $1 \leq R_u^h \leq 136$, while $R_c^g = R_c^g = 50$.

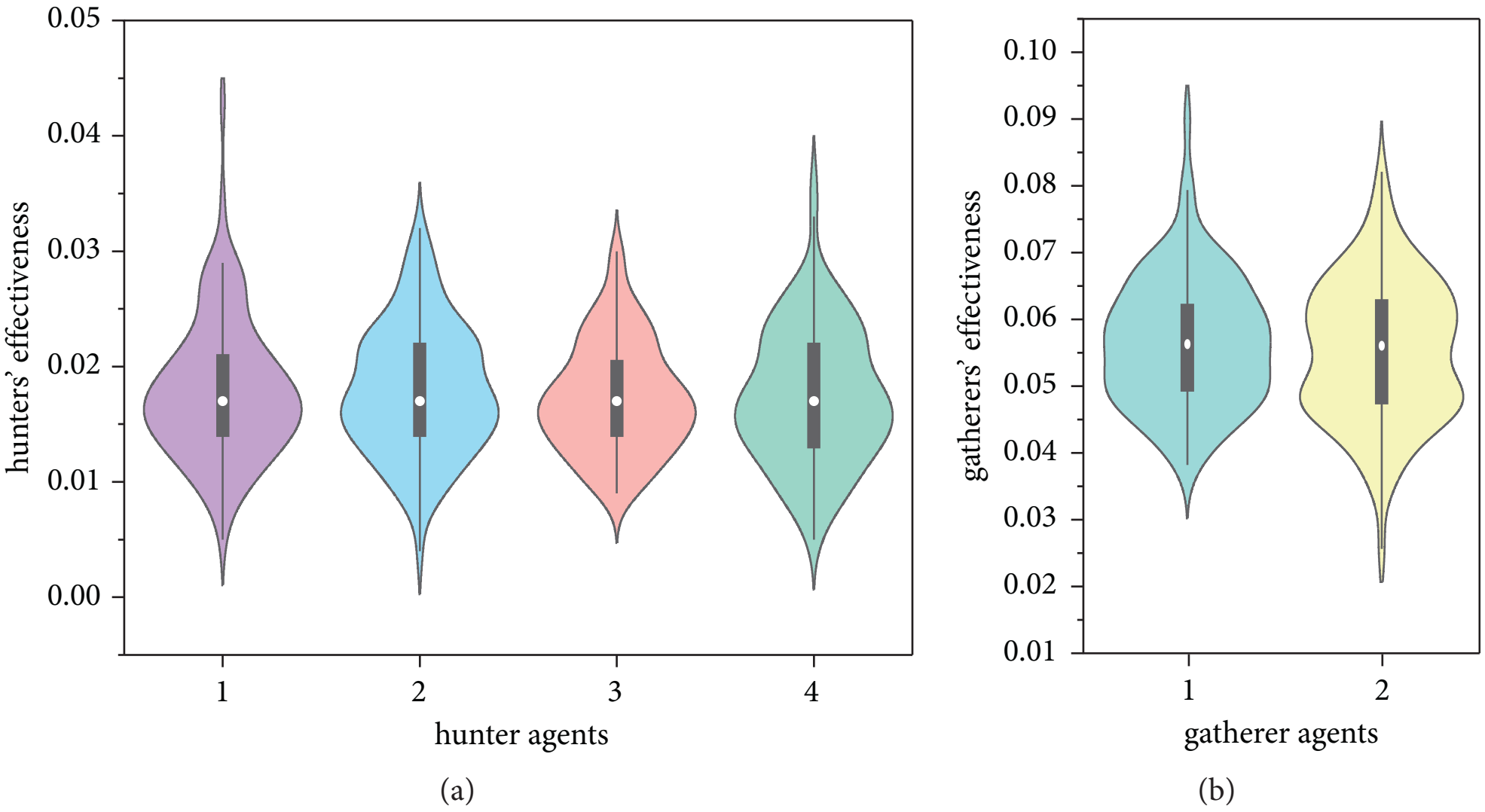

FIGURE 5: Investigating the fairness of the proposed algorithms for (a) hunter agents and (b) gatherer agents.

$\eta_t$ has been calculated for each set of values for $R_c^h$ and $R_u^h$, as illustrated in Figure 6(a). The results show that $\eta_t$ ranges from 0 to 0.012 while changing the value of $R_c^h$ and $R_u^h$ during the whole simulation. As Figure 6(a) displays, the total effectiveness reaches its maximum value, when $R_c^h = 10$ and $R_u^h = 110$. Besides, the slight increase in the total effectiveness in Figure 6(a) for small values of $R_c^h$ and $R_u^h$ can be interpreted as the effect of perpetual running of the simulations. In perpetual mode, known areas get unknown after a certain number of iterations which inevitably favors agents with smaller profit margins. Because of that, although a hunter agent with small values of $R_c^h$ and $R_u^h$ cannot explore distant frontiers to detect tasks, its surrounding explored areas get unknown with a possibility of popping up new tasks. However, the ensemble effect of profit margins of hunters ensures the existence of a maximum for $\eta_t$.

Similar simulations have been run for all values of $R_c^g$ and $R_u^g$, i.e., the profit margins of gatherers, such that $1 \leq R_c^g \leq 136$ and $1 \leq R_u^g \leq 136$, while $R_c^h = R_c^h = 70$. $\eta_t$ has been calculated for each set of values for $R_c^g$ and $R_u^g$, as illustrated in Figure 6(b). The results show that $\eta_t$ranges from 0 to 0.012 while changing the value of $R_c^h$ and $R_u^h$ during the whole simulation. According to the results, the total effectiveness reaches its maximum value when $R_c^h = 30$ and $R_u^h = 40$.

The main conclusion to be drawn is that the introduced profit margin parameters for both types of agents have distinct effect on the total effectiveness and there exists a maximum value for $\eta_t$. Moreover, according to the proposed methodology for both types of agents, when $R_c$ decreases, the agent becomes less confident, and when $R_u$ increases, the agent becomes less conservative. In this regard, for both types of agents, the best strategy to reach the maximum of $\eta_t$ is neither being completely confident nor being fully conservative, but a combination of both leads to the optimum result.

*4.3. Effect of Coordination Factor on the Total Effectiveness.* In this section, we aim to investigate the effect of the coordination factor, introduced in Section 3.4, on the mission's effectiveness. To conduct a comprehensive investigation, we carry out the experiments in three different grid maps with various levels of obstacle complexities: (a) a simple grid map containing two straight barriers, (b) a grid map containing sparse obstacles, and (c) a confined grid map containing narrow corridors and confined rooms, as depicted in Figure 7.

After defining three different grid maps, we ran the algorithm 200 times for different values of $\mu$ in each defined grid map, as illustrated in Figure 8. First, these results show that there is a value for $\mu$ in each grid map that leads to a maximum value of $\overline{\eta}_t$ that denotes the average of $\eta_t$ in 200 tests. Considering that we want to know that how much $\overline{\eta}_t$ increases when coordination factor changes from $\mu = 0$ to $\mu = \mu_{\max}$. In fact, this investigation compares two cases: (1) task planning without any coordination between gatherers and hunters ($\mu = 0$), and (2) task planning with the optimum value of the coordination factor ($\mu = \mu_{\max}$). For that purpose, we applied a paired $T$-test to two of the collected datasets from Figure 8(a). The first dataset contains 200 measures of $\eta_t$ for$\mu = 0$, and the second dataset comprises 200 measures of $\eta_t$ for$\mu = \mu_{\max} = 0.4$. The test has been conducted considering $H_0$: $\overline{\eta}_2 - \overline{\eta}_1 \leq D_0$, $H_1$: $\overline{\eta}_2 - \overline{\eta}_{11} > D_0$, $D_0 = 0.15\overline{\eta}_1$, $n_s = 200$, *do f* = 199, and $\alpha = 0.05$ where $\overline{\eta}_1$ and $\overline{\eta}_2$ denote the average of $\eta_t$ for the first and second datasets, respectively. According to the test result, $p - \text{value} = 0.017$, $t = 2.12$, and $t_{0.05,99} = 1.65$. Since $t > t_{0.05,99}$ and $p - \text{value} < \alpha$, we reject $H_0$. Therefore, the results prove that $\eta_t$ increases more than 15 percent by changing $\mu$from 0 to 0.4.

Likewise, we applied the same statistical analysis on the datasets collected from two other grid maps, i.e., the

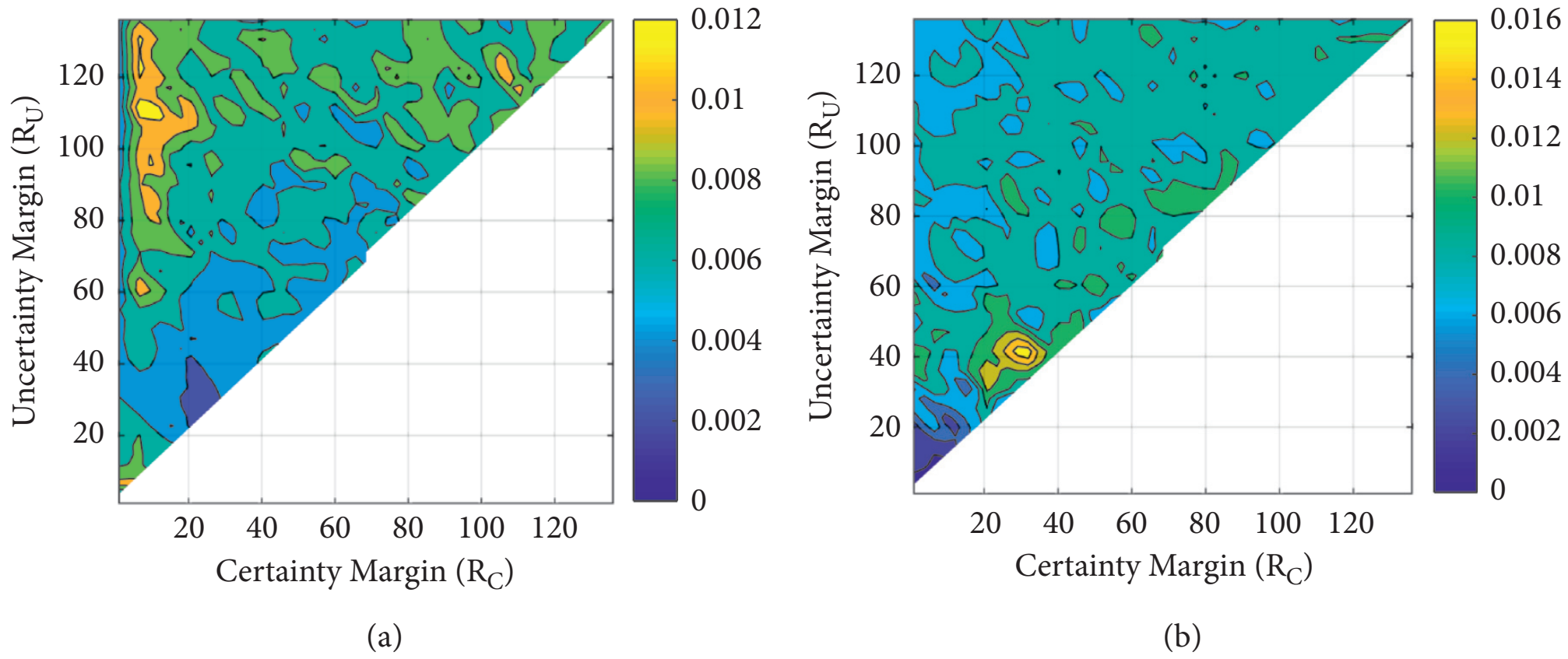

FIGURE 6: Investigating the effect of hunters' and gatherers' profit margins on the total effectiveness: (a) the contour plot of the results for hunters' profit margins, and (b) the contour plot of the results for gatherers' profit margins.

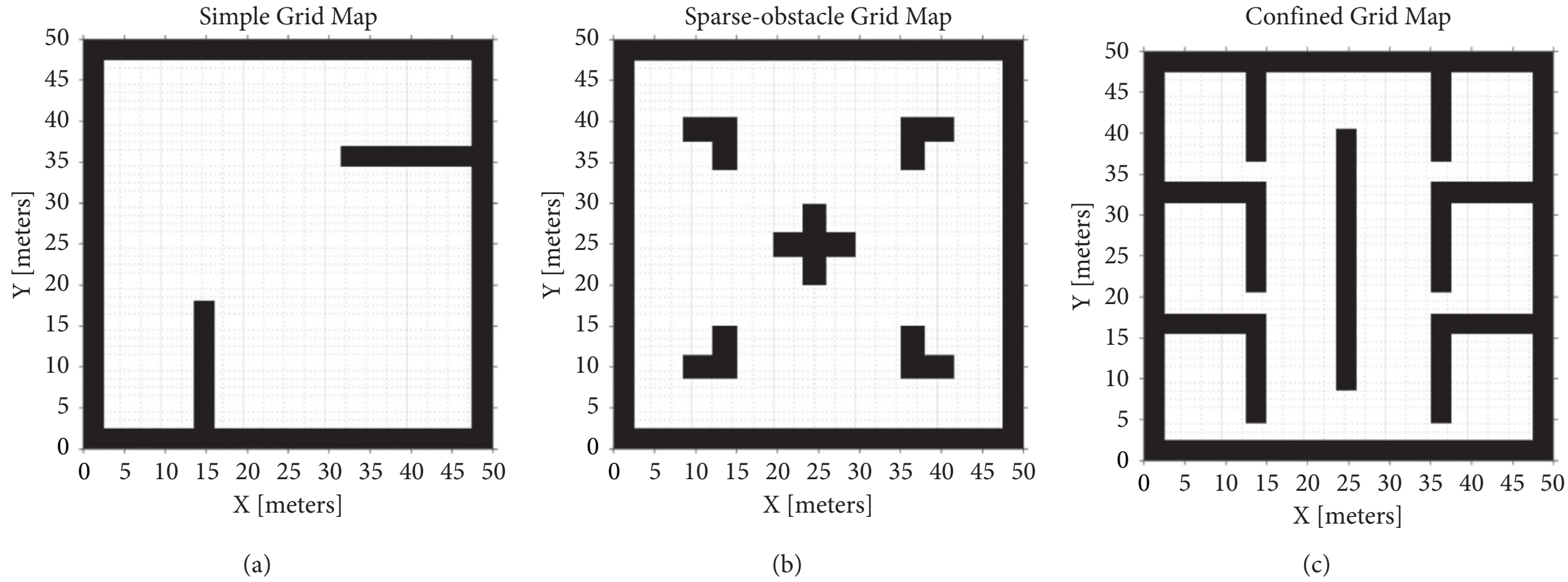

FIGURE 7: Three grid maps with different levels of obstacle complexities: (a) a simple gird map, (b) a grid map with sparse obstacles, and (c) a confined grid map.

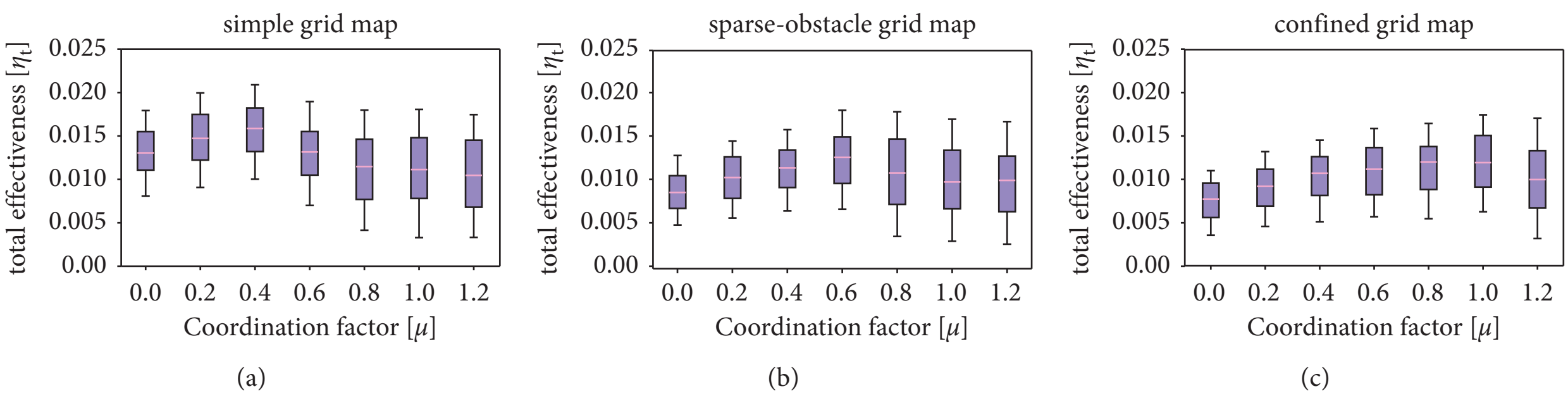

FIGURE 8: Investigating the effect of coordination factor on the planning's total effectiveness in (a) a simple grid map, (b) a grid map with sparse obstacles, and (c) a confined grid map.

environment with sparse obstacles and with confined obstacles, as shown in Figure 8(b) and 8(c). According to the analysis for the environment with sparse obstacles, $\eta_t$ increases by 35 percent when $\mu$ changes from 0 to its optimum value which is 0.6. In addition, the analysis suggests that by changing $\mu$ from 0 to its optimum value for the environment containing confined obstacles, $\eta_t$ increases by 60 percent. Considering the statistical analysis for results in all three environments, there are two insightful implications about the effects of the coordination factor in different environments:

First, the more the complexity of the obstacles in the environment, the more the drop of the total effectiveness of the planning algorithms. On this subject, the total effectiveness without any coordination between hunters and gatherers ($\mu = 0$) drops by 32.4% from an environment with simple obstacles to the one with sparse obstacles. This drop is even more serious, that is 45.7%, by changing the obstacles of the environment from simple obstacles to confined ones. All told, lack of coordination between cooperating agents, i.e., hunters and gathers, is relatively more critical problem when the environment comprises more complex obstacles and confined and narrow corridors.

Secondly, when the environment contains more complex obstacles, it takes higher values of the coordination factor $\mu$to prevent the total effectiveness from dropping significantly. As the results presented in Figure 8 suggested, the optimum value of $\mu$ is 0.4 in an environment with simple obstacles, while the optimum value increases to 0.6 and 1 with sparse and confined obstacles, respectively. All over again, this finding emphasizes the criticality and necessity and criticality of coordination between cooperating agents from complementary teams.

*4.4. Functionality Validation of the Hunter-and-Gatherer Approach.* In this section, we aim to compare the newly developed planning algorithms addressing the exploration and coordination aspects of the hunter-and-gatherer scenario with the benchmark hunter-and-gatherer mission planning introduced in [9]. To that end, we conduct the comparison in three environments containing different configurations of obstacles, as described in Figure 7. 200 tests have been carried out for each solution in each environment, as Figure 9 displays the results. To draw a valid comparison, the mutual parameters in both solutions are set identically as $n_h = 4$, $n_g = 2$, $m_p = 50$, $e_L = e_W = 100$, $\tau_{\max} = 1000$, and $\rho_h/\rho_g = 0.2$. Other specific parameters are set to their optimal values for both solutions.

According to the results, the new approach presented in this paper performs significantly superior to the benchmark hunter-and-gatherer mission planning in all three different environments. To demonstrate this statistically for the environment with simple obstacles, we apply a paired $T$-test to two of the collected datasets, where the first dataset contains 200 measures of $\eta_t$ for the new approach introduced in this work, and the second dataset comprises 200 measures of $\eta_t$ for the benchmark hunter-and-gatherer approach. The statistical analysis suggests that the new approach performs more effective than the benchmark approach by 14 percent, where $n_s = 200$, $do\ f = 199$, $p-\text{value} = 0.026$, $t = 1.95$, and $t_{0.05,99} = 1.66$. By the same token, the new approach outperforms the benchmark solution by 28% and 36% in the environments with sparse and confined obstacles, respectively.

The analysis discussed above indicates that the amount of the improvement is correlated with the complexity of the obstacles in the environment. Having said that, addressing the exploration and coordination of the hunter-and-gatherer scenario is much more critical when the environment comprises obstacles, especially dense obstacles. That is mainly because in the case of the benchmark approach performing in confined environments including dense obstacles, agents are more susceptible to get far away from each other, which eliminates any overlap between agent's profit margins. Consequently, all sorts of negotiations converge to being declined since the distance factor makes all negotiations unprofitable. Although in the new approach the agents still rely on their profit margins to make the decisions, the coordination factor between hunters and gatherers facilitates keeping reasonable overlap in their profit margins. This leads to a relatively much more effective and efficient performance in environment with the presence of obstacles, compared to the benchmark method.

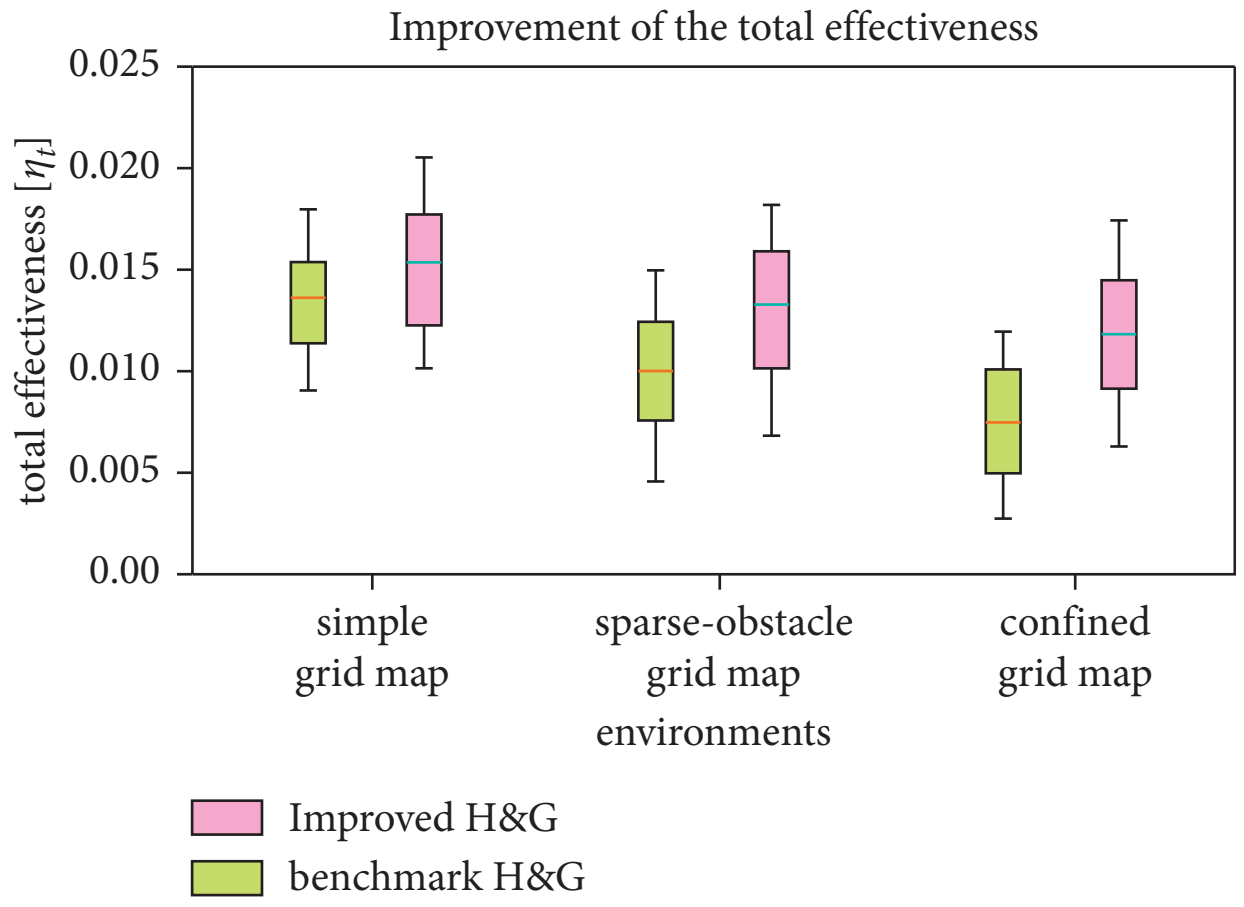

Figure 9: Comparing the proposed approach with the benchmark hunter-and-gatherer approach in three different environments.

To further investigate the performance of the proposed planning method, we incorporated different multirobot exploration algorithms into the developed platform and measured the total effectiveness of the planning, as defined in equation (14). In other words, we maintained the proposed algorithm for gatherers and investigated the performance of the proposed method with different search algorithms for hunters, including the one proposed in this paper called EG-based frontier selection. This experiment has been conducted on the confided grid map introduced in Figure 7(c), and for each exploration algorithm, the experiment has been repeated 20 times for each value of $n_t$. As demonstrated in Figure 10, the random walk multiagent exploration algorithm [41] performed ineffectual compared to other algorithms. This is mainly because the random walk search algorithm neglects the distance to the frontiers and disregards the locations of other hunters in the environment. By considering the distance to candidate frontiers, multiagent frontier-based exploration algorithm [42] enhances the total effectiveness of the planning by 93.4% compared to the results with the random walk algorithm. The most optimal performance of the frontier-based exploration algorithm occurs with eight hunters in the confined environment, while the random walk algorithm requires ten hunters for the very same environment to reach the optimal performance.

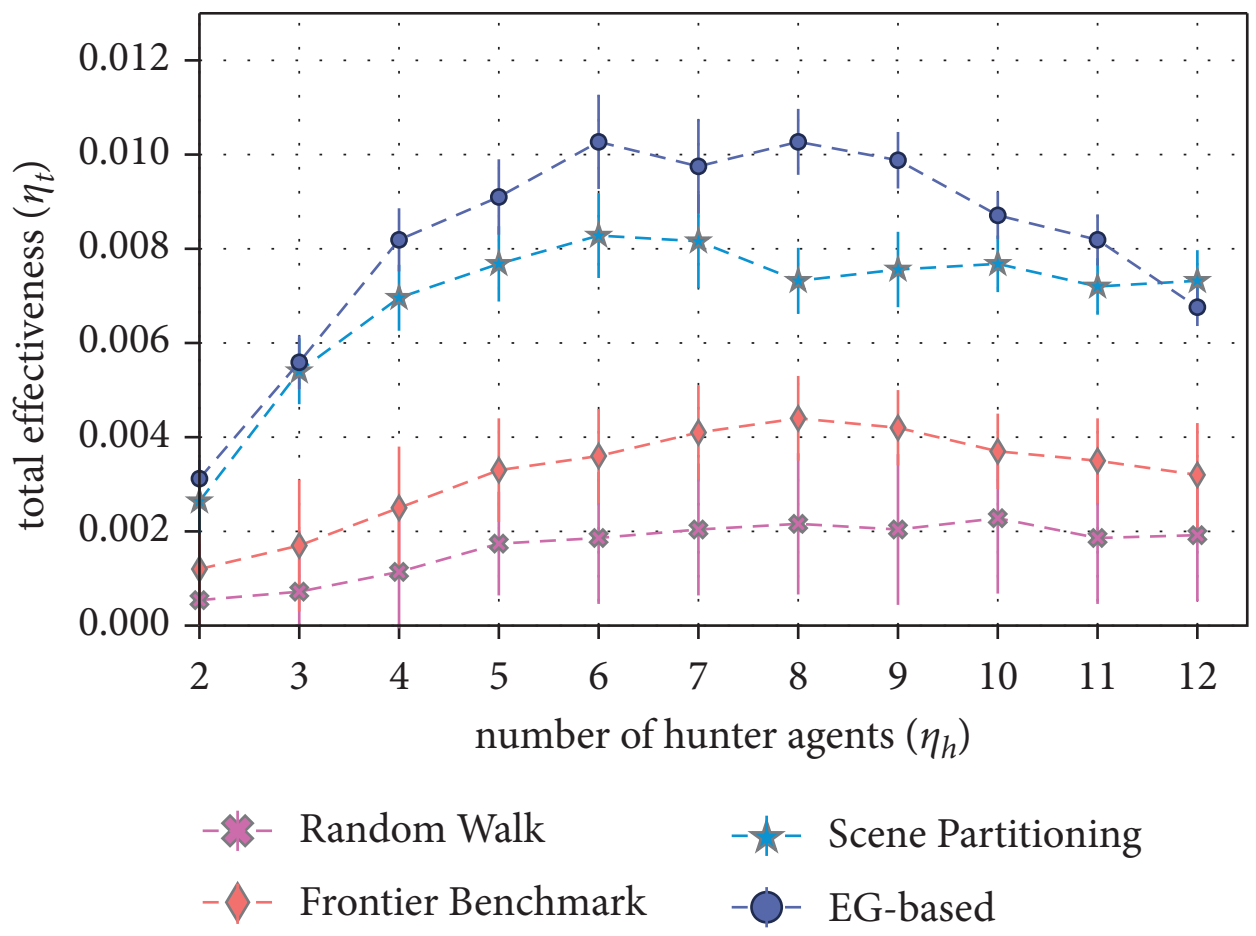

FIGURE 10: Analyzing the total effectiveness of the proposed approach with various multirobot exploration algorithms including the EG-based method introduced in this work.

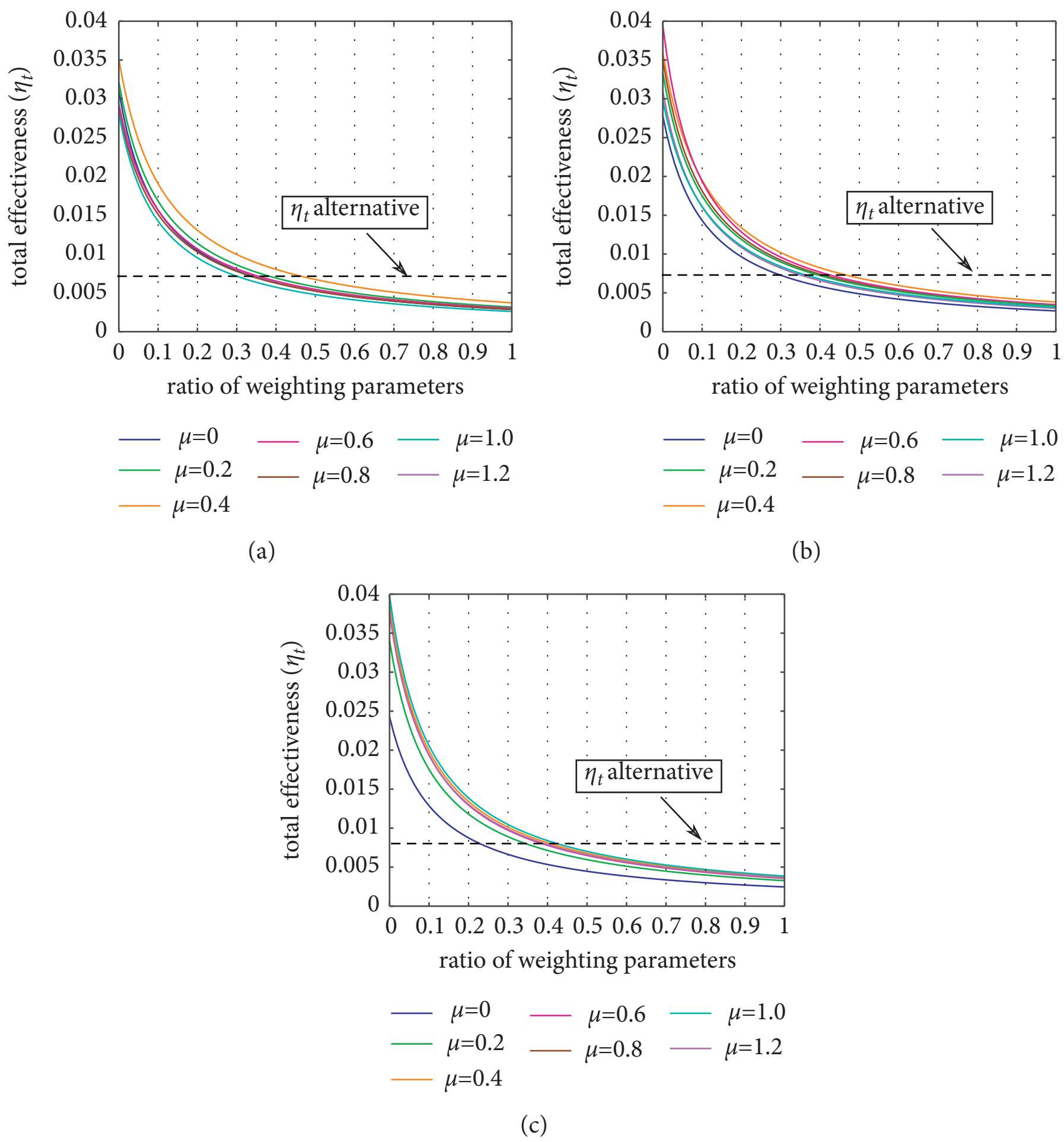

FIGURE 11: Comparing the hunter-and-gatherer approach with an alternative approach where each agent does both exploration and completion together.

Alternatively, the scene partitioning multiagent exploration algorithm [20] performs drastically superior to the frontier-based algorithm and enhances the total effectiveness by 115.2% compared to the results with the frontier-based algorithm. The partitioning mechanism of this algorithm decreases the conflict and overlapping between explorers which leads to relatively more efficient performance. Furthermore, the scene partitioning algorithm requires six hunters to perform optimally. Finally, the EG-based multirobot exploration algorithm presented in this paper outperforms all three algorithms. Although the EG-based exploration algorithm similarly requires 6 hunters to perform optimally, it enhances the scene partitioning results by 17.8%. In fact, EG-based exploration algorithm facilitates a dynamic and temporal partitioning which provides more flexibility and controlled freedom for agents to decide and plan optimally.

*4.5. Functionality Validation of the Hunter-and-Gatherer Approach.* To validate the functionality of the hunter-and-gatherer approach, we compared the proposed approach with an alternative approach in which there is only one type of agent doing both exploration and completion of tasks together. The goal of this comparison is to answer two critical questions: (1) is hunter-and-gatherer scheme more economic than the explained alternative approach? and (2) what criterion needs to be satisfied for the hunter-and-gatherer scheme to be relatively economic?

The hunter-and-gatherer approach fundamentally differs from the alternative approach in employing two types of agile (hunters) and dexterous (gatherers) agents. In contrast to the alternative approach, each task takes two agents, that is, a pair of hunter-and-gatherer agents, to be completed in the hunter-and-gatherer scheme. This makes the ratio of the weighting parameters $\rho_h/\rho_g$ an imperative factor to conduct the study. Hence, the criterion judging the functionality of the hunter-and-gatherer scheme should be expressed with respect to the ratio of the weighting parameters. To that end, we ran the algorithm for different values of $\mu$in each map and compared its total effectiveness with the total effectiveness of the alternative approach in three previously defined environments, as shown in Figure 11. According to the results, it is economic to employ the proposed hunter-and-gatherer approach for dynamic ST-MR-TA : SP problems if and only if we utilize hunter-and-gatherer agents that satisfy $\rho_h/\rho_g < 0.45$ approximately. In other words, if the cost of accomplishment of a hunter agent for a certain task is less than 45 percent of the gather's accomplishment cost for the same task, then employing the hunter-and-gatherer scheme, instead of the alternative approach, is relatively economic.

## 5. Conclusion

This paper addresses the multirobot task allocation problem in dynamic environments and focuses on the exploration and coordination aspects of the hunter-and-gatherer scheme. On this line of thought, we proposed an innovative decision-making mechanism based on the novel notion of EG, which measures the density of available information in the surrounding of a potential job (task/frontier). We demonstrated that applying the EG-based decision-making mechanism on hunters and gatherers to address exploration and task allocation aspects of the problem improves the performance of the hunter-and-gatherer scheme significantly compared to the authors' previous negotiation-based solution. We found that the significance of the improvement is correlated with the complexity of obstacles in the environment. Besides, this work proposed an EG-based coordination algorithm for gatherers, which led to a momentous increase in the planning's effectiveness. Statistical analysis on the simulations' results suggests that there is an optimum value for the coordination factor which maximizes the planning's total effectiveness. Likewise, we found that the optimum value of the coordination factor varies for environments with different densities and difficulties of obstacles. Collectively, the higher the complexity and difficulty of the obstacles in an unknown environment, the more the improving effect of the proposed method on the planning's total effectiveness. Moreover, we showed that employing two complementary teams of hunters and gatherers can effectually improve the total effectiveness of the task allocation in a mission. However, this is only true when the defined judging criteria, associated with the ratio of the weighting parameters, is adequately satisfied. Practically speaking, the affordability criteria comparing relative costs of each type of agent are straightforwardly satisfiable, as the USAR case exemplifies a real-world problem where the relative accomplishment costs of hunters (small UAVs) and gatherers (heavy-duty UGVs) satisfy the defined criteria.

Future research should define the cost function more comprehensively by considering the communication burden between agents. This provides more realistic settings to evaluate the efficacy of solutions utilizing communication, such as the hunter-and-gatherer scheme. Besides, further research is needed to confirm the functionality of the hunter-and-gatherer scheme in practice by carrying out a multirobot test bench to study the complexities and limitations of the developed theories in the context of the hunter-and-gatherer approach. Future works can also consider optimizing the control parameters introduced in this work utilizing a wide variety of optimization or learning methods.

## Data Availability

The data and source code used to support the findings of this study are available from the corresponding author upon request. This paper has been archived at Cornell University Library [43].

## Conflicts of Interest

The authors declare that they have no conflicts of interest.

## Acknowledgments

This research was supported by Lamar University via internal grants.

## References

[1] S. A. Nelke, S. Okamoto, and R. Zivan, "Market clearing-based dynamic multi-agent task allocation," *ACM Transactions on Intelligent Systems and Technology*, vol. 11, no. 1, pp. 1–25, 2020.

[2] M. J. Matarić, G. S. Sukhatme, and E. H. Østergaard, "Multi-robot task allocation in uncertain environments," *Autonomous Robots*, vol. 14, no. 2, pp. 255–263, 2003.

[3] W. Burgard, M. Moors, C. Stachniss, and F. E. Schneider, "Coordinated multi-robot exploration," *IEEE Transactions on Robotics*, vol. 21, no. 3, pp. 376–386, 2005.

[4] W. Sheng, Q. Yang, J. Tan, and N. Xi, "Distributed multi-robot coordination in area exploration," *Robotics and Autonomous Systems*, vol. 54, no. 12, pp. 945–955, 2006.

[5] E. Asali, M. Valipour, N. Zare, A. Afshar, M. Katebzadeh, and G. H. Dastghaibyfard, "Using Machine Learning Approaches to Detect Opponent Formation," in *Proceedings of the 2016 Artificial Intelligence and Robotics (IRANOPEN)*, pp. 140–144, Qazvin, Iran, April 2016.

[6] H. Haeri, K. Jerath, and J. Leachman, "Thermodynamics-inspired macroscopic states of bounded swarms," *ASME Letters in Dynamic System Control*, vol. 1, no. 1, 2021.

[7] H. Haeri, K. Jerath, and J. Leachman, "Thermodynamics-inspired modeling of macroscopic swarm states," in *Proceedings of the AMSE 2019 Dynamic Systems and Control Conference*, Park City, UT, USA, October 2019.

[8] F. Ye, J. Chen, Q. Sun, Y. Tian, and T. Jiang, "Decentralized task allocation for heterogeneous multi-UAV system with task coupling constraints," *The Journal of Supercomputing*, vol. 77, no. 1, pp. 111–132, 2021.

[9] M. Dadvar, S. Moazami, H. R. Myler, and H. Zargarzadeh, "Multiagent task allocation in complementary teams: a hunter-and-gatherer approach," *Complexity*, vol. 2020, Article ID 1752571, 2020.

[10] M. Islam, M. Dadvar, and H. Zargarzadeh, "A dynamic territorializing approach for multiagent task allocation," *Complexity*, vol. 2020, Article ID 8141726, 2020.

[11] B. P. Gerkey and M. J. Matarić, "A formal analysis and taxonomy of task allocation in multi-robot systems," *The International Journal of Robotics Research*, vol. 23, no. 9, pp. 939–954, 2004.

[12] H.-L. Han-Lim Choi, L. Brunet, and J. P. How, "Consensus-based decentralized auctions for robust task allocation," *IEEE Transactions on Robotics*, vol. 25, no. 4, pp. 912–926, 2009.

[13] A. Jevtić and A. Gutiérrez, "Distributed bees algorithm parameters optimization for a cost efficient target allocation in swarms of robots," *Sensors*, vol. 11, no. 11, pp. 10880–10893, 2011.

[14] S. Liemhetcharat and M. Veloso, "Weighted synergy graphs for effective team formation with heterogeneous ad hoc agents," *Artificial Intelligence*, vol. 208, pp. 41–65, 2014.

[15] B. Yamauchi, "A frontier-based approach for autonomous exploration," in *Proceedings 1997 IEEE International Symposium on Computational Intelligence in Robotics and Automation CIRA'97. "Towards New Computational Principles for Robotics and Automation"*, pp. 146–151, Monterey, CA, USA, July 1997.

[16] B. Yamauchi, "Frontier-based exploration using multiple robots," in *Proceedings of the Second International Conference on Autonomous Agents*, pp. 47–53, Minneapolis, MN, USA, May 1998.

[17] R. Zlot, A. Stentz, M. B. Dias, and S. Thayer, "Multi-robot exploration controlled by a market economy," in *Proceedings 2002 IEEE international conference on robotics and automation (Cat. No. 02CH37292)*, vol. 3, pp. 3016–3023, Washington, DC, USA, May 2002.

[18] S. Bhattacharya, R. Ghrist, and V. Kumar, "Multi-robot coverage and exploration on Riemannian manifolds with boundaries," *The International Journal of Robotics Research*, vol. 33, no. 1, pp. 113–137, 2014.

[19] S. Bhattacharya, N. Michael, and V. Kumar, "Distributed coverage and exploration in unknown non-convex environments," *Distributed Autonomous Robotic Systems*, Springer, Berlin, Germany, pp. 61–75, 2013.

[20] J. J. Lopez-Perez, U. H. Hernandez-Belmonte, J.-P. Ramirez-Paredes, M. A. Contreras-Cruz, and V. Ayala-Ramirez, "Distributed multirobot exploration based on scene partitioning and frontier selection," *Mathematical Problems in Engineering*, vol. 2018, Article ID 2373642, 2018.

[21] A. Prorok, M. A. Hsieh, and V. Kumar, "The impact of diversity on optimal control policies for heterogeneous robot swarms," *IEEE Transactions on Robotics*, vol. 33, no. 2, pp. 346–358, 2017.

[22] D. Wu, G. Zeng, L. Meng, W. Zhou, and L. Li, "Gini coefficient-based task allocation for multi-robot systems with limited energy resources," *IEEE/CAA Journal of Automatica Sinica*, vol. 5, no. 1, pp. 155–168, 2017.

[23] J. Faigl, O. Simonin, and F. Charpillet, "Comparison of task-allocation algorithms in frontier-based multi-robot exploration," in *Proceedings of the European Conference on Multi-Agent Systems*, pp. 101–110, Prague, Czech Republic, December 2014.

[24] E. Nunes, M. Manner, H. Mitiche, and M. Gini, "A taxonomy for task allocation problems with temporal and ordering constraints," *Robotics and Autonomous Systems*, vol. 90, pp. 55–70, 2017.

[25] S. Habibian, M. Dadvar, B. Peykari et al., "Design and implementation of a maxi-sized mobile robot (Karo) for rescue missions," *ROBOMECH Journal*, vol. 8, no. 1, pp. 1–33, 2021.

[26] S. Poonguzhali and T. Gomathi, "Design and implementation of ploughing and seeding of agriculture robot using IOT," *Soft Computing Techniques and Applications*, Springer, Berlin, Germany, pp. 643–650, 2021.

[27] C. A. My, S. S. Makhanov, N. A. Van, and V. M. Duc, "Modeling and computation of real-time applied torques and non-holonomic constraint forces/moment, and optimal design of wheels for an autonomous security robot tracking a moving target," *Mathematics and Computers in Simulation*, vol. 170, pp. 300–315, 2020.

[28] Q. Lu, G. M. Fricke, J. C. Ericksen, and M. E. Moses, "Swarm foraging review: closing the gap between proof and practice," *Current Robotics Reports*, vol. 1, pp. 1–11, 2020.

[29] S. Saju Sankar and S. S. Vinod Chandra, "A multi-agent ant colony optimization algorithm for effective vehicular traffic management," in *Proceedings of the ICSI 2020 International Conference on Swarm Intelligence*, pp. 640–647, Belgrade, Serbia, July 2020.

[30] C. Hahn, F. Ritz, P. Wikidal, T. Phan, T. Gabor, and C. Linnhoff-Popien, "Foraging swarms using multi-agent reinforcement learning," in *Proceedings of the ALIFE 2020 Artificial Life Conference Proceedings*, pp. 333–340, Montréal, Canada, July 2020.

[31] W. Lee, N. Vaughan, and D. Kim, "Task allocation into a foraging task with a series of subtasks in swarm robotic system," *IEEE Access*, vol. 8, pp. 107549–107561, 2020.

[32] J. T. Isaacs, N. Dolan-Stern, M. Getzinger, E. Warner, A. Venegas, and A. Sanchez, "Central place foraging: Delivery lanes, recruitment and site fidelity," in *Proceedings of the 2020*

*IEEE International Conference on Autonomous Robot Systems and Competitions (ICARSC)*, pp. 319–324, Ponta Delgada, Portugal, April 2020.

[33] T. Deemyad, N. Hassanzadeh, and A. Perez-Gracia, "Coupling mechanisms for multi-fingered robotic hands with skew axes," in *Proceedings of the IFToMM Symposium on Mechanism Design for Robotics*, pp. 344–352, Udine, Italy, Auguest 2018.

[34] T. Deemyad, O. Heidari, and A. Perez-Gracia, "Singularity design for RRSS mechanisms," in *Proceedings of the USCToMM Symposium on Mechanical Systems and Robotics*, pp. 287–297, Rapid City, SD, USA, May 2020.

[35] R. R. Murphy, *Search and Rescue Robotics*, pp. 1151–1173, Springer, Berlin, Germany, 2007.

[36] O. Goldreich, "Finding the shortest move-sequence in the graph-generalized 15-puzzle is NP-hard," *Studies in Complexity and Cryptography. Miscellanea on the Interplay between Randomness and Computation*, Springer, Berlin, Germany, pp. 1–5, 2011.

[37] D. Ratner and M. Warmuth, "The ($n^2$–1)-puzzle and related relocation problems," *Journal of Symbolic Computation*, vol. 10, no. 2, pp. 111–137, 1990.

[38] H. Aziz, H. Chan, Á. Cseh, B. Li, F. Ramezani, and C. Wang, *Multi-Robot Task Allocation-Complexity and Approximation*, in *AAMAS'21: Proceedings of the 20th International Conference on Autonomous Agents and MultiAgent Systems*, Virtual Event UK, May 2021.

[39] B. Kim, C. M. Kang, J. Kim, S. H. Lee, C. C. Chung, and J. W. Choi, "Probabilistic vehicle trajectory prediction over occupancy grid map via recurrent neural network," in *Proceedings of the 2017 IEEE 20th International Conference on Intelligent Transportation Systems (ITSC)*, pp. 399–404, Yokohama, Japan, October 2017.

[40] D. C. Montgomery, G. C. Runger, and N. F. Hubele, *Engineering Statistics*, John Wiley & Sons, Hoboken, NJ, USA, 2009.

[41] B. Pang, Y. Song, C. Zhang, and R. Yang, "Effect of random walk methods on searching efficiency in swarm robots for area exploration," *Applied Intelligence*, vol. 51, no. 7, pp. 5189–5199, 2021.

[42] Y. Wang, A. Liang, and H. Guan, "Frontier-based multi-robot map exploration using particle swarm optimization," in *Proceedings of the 2011 IEEE Symposium on Swarm Intelligence*, pp. 1–6, Paris, France, April 2011.

[43] M. Dadvar, S. Moazami, H. R. Myler, and H. Zargarzadeh, "Exploration and coordination of complementary multi-robot teams in a hunter and gatherer scenario," 2019, http://arxiv.org/abs/1912.07521.